\documentclass[11pt]{amsart}
\usepackage{amsmath,amsfonts,latexsym,amssymb,amscd, graphicx,enumerate}
\usepackage[normalem]{ulem}
\usepackage[usenames]{color}

\usepackage{graphicx,psfrag,epsfig}
\usepackage{amssymb,amsmath,amscd,amsthm,mathrsfs}
\usepackage{graphicx,psfrag,epsfig,xcolor}
\usepackage{bm,enumerate}
\usepackage{times}
\usepackage{graphicx}
\usepackage{mathtools}
\usepackage{tikz}
\usepackage{microtype}
\usepackage{soul}
\usepackage{amsfonts}
\usepackage{color}
\usepackage{comment}
\usepackage[normalem]{ulem}

\usepackage[makeroom]{cancel}

\usetikzlibrary{decorations.markings}

\usepackage{hyperref}% Put this last
\makeatletter
\@ifpackagelater{hyperref}{2012/05/28}{
  \definecolor{link1}{Hsb}{240,1,.75}
  \definecolor{link2}{Hsb}{240,1,.5}
  \hypersetup{final,colorlinks,allcolors=link1,citecolor=link2}
}{
  \hypersetup{final}
}

\newcommand{\visc}{\nu}

\newcommand{\bpf}[1][Proof]{{\noindent {\sc #1: }}}
\newcommand{\epf}{{{\hfill $\Box$ \smallskip}}}
\newcommand{\ONE}{{\mathbf{1}}}

\newcommand{\N}{{\mathbb N}}

\newcommand{\Fc}{\mathcal{F}}

\newcommand{\Pp}{\mathsf{P}}
\newcommand{\Z}{{\mathbb Z}}

\newcommand{\T}{{\mathbb T}}

\newcommand{\E}{\mathsf{E}}

\newcommand{\R}{{\mathbb R}}

\newcommand{\X}{\mathbb{X}}
\newcommand{\Xc}{\mathcal{X}}

\newcommand{\Lc}{\mathcal{L}}
\newcommand{\argmax}{\mathop{\mathrm{argmax}}}
\newcommand{\argmin}{\mathop{\mathrm{argmin}}}
\newcommand{\eff}{\mathrm{eff}}
\newcommand{\Ai}{\mathrm{Ai}}

\newcommand{\Qq}{\mathsf{Q}}
\newcommand{\Ww}{\mathsf{W}}

\newtheorem{theorem}{Theorem}
\newtheorem{conjecture}{Conjecture}
\newtheorem*{statement*}{Statement}

\title[Random Hamilton--Jacobi equations]{On global solutions of the random Hamilton--Jacobi equations and the KPZ problem}

\author{Yuri Bakhtin}\thanks{Courant Institute of Mathematical Sciences, New
  York University, 251 Mercer St, New York, NY 10012 USA;  {\it email: bakhtin@cims.nyu.edu}}
\author{Konstantin Khanin}
\thanks{Department of Mathematics, University of Toronto, Bahen
  Centre, 40 St.~George St, Toronto, ON M5S 2E4 Canada; Institute for Information Transmission Problems (Kharkevich
Institute), Russian Academy of Sciences
Bolshoy Karetny per.19, build.1, Moscow 127051 Russia;
{\it email:khanin@math.toronto.edu}}

\begin{document}

\begin{abstract}
In this paper, we discuss possible qualitative approaches to the problem
of KPZ universality. Throughout the paper, our point of view is based on
the geometrical and dynamical properties of minimisers
and shocks forming interlacing tree-like structures.
We believe that the KPZ universality can be explained in terms
of statistics of these structures evolving in time.
The paper is focussed  on the setting of the random Hamilton--Jacobi equations.
We formulate several conjectures concerning global solutions
and discuss how their
properties are connected to the KPZ scalings
in dimension 1+1.
In the case of general viscous Hamilton--Jacobi equations with non-quadratic Hamiltonians, we
define generalised directed polymers. We expect that their behaviour is similar
to the behaviour of classical directed polymers, and present arguments in favour
of this conjecture. We also define a new renormalisation transformation defined
in purely geometrical terms and discuss conjectural properties of the
corresponding
fixed points. Most of our conjectures are widely open, and supported by only
partial rigorous results for particular models.
\end{abstract}
\maketitle
\section{Introduction}
\label{sec:intro}
The problem of the KPZ phenomenon and universality has been one of the most active directions in statistical physics in the last decade, see~
\cite{Alberts-Khanin-Quastel:MR3189070},
\cite{Amir-Corwin-Quastel:CPA20347},
\cite{Baik-Deift-Johansson:MR1682248},
\cite{Balazs-Cator-Seppalainen:MR2268539},
\cite{Borodin-Corwin:MR3152785},
\cite{Borodin-Corwin-Ferrari:MR3207195},
\cite{Borodin-Corwin-Ferrari-Vetho:MR3366125},
\cite{Borodin-Ferrari:MR2438811},
\cite{Borodin-Ferrari-Sasamoto:MR2430639},
\cite{Borodin-Gorin:MR3526828},
\cite{Calabrese-Le-Doussal-Rosso:0295-5075-90-2-20002},
\cite{Calabrese-Le-Doussal:PhysRevLett.106.250603},
\cite{Cator-Groeneboom:MR2257647},
\cite{CaPi-ptrf},
\cite{Corwin:MR2930377},
\cite{Corwin-Quastel-Remenik:MR3373642},
\cite{Dotsenko:0295-5075-90-2-20003},
\cite{Gubinelli-Perkowsky:MR3592748},
\cite{Hairer:MR3274562},
\cite{Hairer:MR3071506},
\cite{Le-Doussal-Calabrese:1742-5468-2012-06-P06001},
\cite{Moreno-Remenik-Quastel:MR3010188},
\cite{Prahofer-Spohn:MR1933446},
\cite{Quastel-Spohn:MR3373647},
\cite{Sasamoto-Spohn:MR2628936},
\cite{Sasamoto-Spohn:PhysRevLett.104.230602},
\cite{Seppalainen:MR2917766},
\cite{Tracy-Widom:MR1385083},
and  multiple other contributions.
 A fascinating feature of the problem is a combination of two factors: exact solvability and universality. On the one hand, one
can write exact formulas for the limiting objects of some particular models. On the other hand, these formulas are supposed to describe the limiting behavior of a huge class of systems that are not integrable. The universality is so global that in a certain sense we do not know how far it stretches. One can say that any large scale ``directed'' variational problem in 2-dimensional disordered media is expected to belong to the KPZ universality class. In this paper, we concentrate not on exact solutions but rather on the universality phenomenon. 

Another remarkable aspect of the problem is the fact that it is tightly connected to diverse areas of mathematics and theoretical physics: PDEs, stochastic analysis, dynamical systems, statistical mechanics, random matrix theory, stochastic geometry, representation theory, to name a few.

We want to emphasize the connection with the problem of global solutions to random  Hamilton--Jacobi  equations. We consider both cases --- classical  Hamilton--Jacobi  equations that correspond to the study of minimisers of Lagrangian action also called geodesics and their parabolic regularizations where action minimisers are replaced by directed polymers.  In our approach, geometrical properties of action minimisers and polymers play an important role. We believe that these geometrical properties make random  Hamilton--Jacobi  equations a better playground than random matrices where no geometric aspects are currently known. We should warn the reader that the number of rigorous results in the area is rather limited. In most cases we provide explanations and conjectures rather than theorems. We believe that many problems that we discuss can me attacked mathematically. Others are more challenging. However, we see value in presenting the general picture and the set of ideas describing the phenomena.

The structure of the paper is the following. In Section~\ref{sec:HJ-eq}, we define random  Hamilton--Jacobi  equations in any dimension and discuss their global solutions. We consider the general case of convex Hamiltonians rather than only quadratic ones. 

To discuss properties of solutions, we introduce the notion of directed polymer that generalizes the usual one defined through the Hopf--Cole transformation and Feynman--Kac formula in the quadratic Hamiltonian case. Such generalized polymers are discussed in Section~\ref{sec:HJ-polymers}. We postpone 
the discussion of 
the equivalence of two notions of polymers
in the quadratic Hamiltonian case to Section~\ref{sec:equiv-of-poly-for-Burgers} playing the role of an appendix.

In Section~\ref{sec:shape}, we discuss the phenomenology of KPZ scaling exponents. We introduce the notion of shape function and demonstrate how its strong convexity property is related to $1:2:3$ KPZ scalings  in dimension $1+1$. 

In Section~\ref{sec:min-shocks-1d}, we discuss the properties of minimisers and shocks in the 1D case.
We present rigorous results in the compact case of periodic forcing potentials,
and discuss how the behaviour of minimisers and shocks changes in the non-compact setting.
We also discuss hyperbolic properties of minimisers playing an important role in the picture.

In Section~\ref{sec:renorm-point}, we discuss point fields that play the role of structural backbones of the global geometry of minimzers and directed polymers. These point fields correspond to locations of concentration of minimisers/polymers and shocks separating the domains of attraction to those locations. We then define the renormalisation transformation acting on these point fields  and formulate several conjectures related to fixed points of this transformation and their stability. 
The transformation is defined in purely geometrical terms without involving action values. This is a reflection of monotonicity which is present only in dimension~1. 
The scheme we discuss is more general than the KPZ phenomenon, with a 1-parameter family of fixed points, where the parameter is the exponent describing the rate of decay of density of the point field with time. 
We conjecture that the statistics of these fixed points is universal and reflects only the properties of interlacing between concentration points and shocks.

The KPZ case corresponds to the density decay rate exponent being equal to~$2/3$. Another special value is $1/2$ which is related to stochastic flows of diffeomorphisms and coalescent Brownian motion. This is the simplest instance of the fixed point, and we discuss it in Section~\ref{sec:colescing-BM}. We also discuss the correlation functions of the fixed point which have been shown to exhibit Pfaffian structure. It is natural to ask whether a similar Pfaffian property may hold for other density exponents.

Another renormalization scheme based on Airy sheet is discussed in Section~\ref{sec:renorm-airy}.

We finish with concluding remarks in Section~\ref{sec:concluding}.

{\bf Acknowledgements.} Large parts of this paper were written in  KITP in Santa Barbara and CIRM in Luminy,
and we are grateful to these centers for hospitality. The work of Yuri Bakhtin was partially supported by NSF through grant DMS-1460595.
The work of Konstantin Khanin was supported by NSERC Discovery Grant RGPIN 328565.

\section{Hamilton--Jacobi equations}
\label{sec:HJ-eq}

We will consider the following randomly forced  Hamilton--Jacobi  equation:

\begin{equation}
 \label{eq:general-Hamilton--Jacobi}
 \partial_t \Phi +  H(\nabla \Phi)=\visc\Delta \Phi - F,  
\end{equation}
Here $\Phi=\Phi^\nu(t,x)$, $t\in\R$, $x\in\R^d$, is a scalar function.
The Hamiltonian $H=H(p)$ is assumed to be a convex function of momentum $p=\nabla\Phi\in\R^d$;
$F=F(t,x)$ is the  external force  potential, i.e.,
$f(t,x)=-\nabla F(t,x)$ is the external force. The viscosity parameter $\nu\ge 0$. We shall consider both viscous case 
where $\nu>0$ and the inviscid case where $\nu=0$. In the former case, solutions are smooth 
in space while in the latter, they 
typically devolop shocks, i.e., gradient discontinuities. Although there are many weak solutions with discontinuities, we will consider only the special one called a viscosity solution that can be defined as the limit
of smooth solutions $\Phi^\nu$ as $\nu\downarrow 0$, but better understood via the Lax--Oleinik variational principle described below.

In this system, the randomness comes into play via the forcing that is assumed to be random, $F(t,x)=F_\omega(t,x)$,
where $\omega$ is an element of a probability space $(\Omega,\Fc,\Pp)$.  We will assume that $F$ is space-time translation invariant
and, without loss of generality, represented via a flow of time shifts $(\theta^t)_{t\in\R}$ that are $\Pp$-preserving transformations of $\Omega$
such that $F_\omega(t,\cdot)=F_{\theta^t\omega}(0,\cdot)$. We will also assume that $F$ has smooth in $x$ realizations  with fast spatial decay of correlations but may be white in time. This setup does not create any difficulty in mathematical treatment of the model, and the solutions are well-defined,
see~\cite{ekms:MR1779561}, \cite{Gomes-Iturriaga-Khanin:MR2241814}.

Clearly, if $\Phi$ is a solution of~\eqref{eq:general-Hamilton--Jacobi}, then, for any constant $c\in\R$, $\Phi+c$ is also a solution. We often do not distinguish solutions that differ by a constant. Moreover, often we take the point of view that only spatial increments of potentials $\Phi$
are important and work in the space of potentials modulo additive constants.

We will often use the random Hamiltonian $H_\omega(p,x,t)$ that includes the random potential:
\[
 H_\omega(p,x,t)=H(p)+F_\omega(t,x),
\]
and the random Lagrangian that is defined as the Legendre transform of the random Hamiltonian:

\[
L_\omega(v,x,t)=\sup_{p}( p\cdot v-H_\omega(p,x,t))=L(v)-F_\omega(t,x),
\]
where
\[
L(v)=\sup_{p}( p\cdot v-H(p)). 
\]
Here $p$ and $v$ form a pair of dual variables with $p=\nabla_v L(v)$ and $v=\nabla_p H(p)$.
We will always assume that the Lagrangian $L(v)$ grows superlinearly as $|v|\to\infty$.

Let us describe the Lax--Oleinik variational principle for the solution of the initial-value problem for the inviscid
 Hamilton--Jacobi  equation, with a continuous initial condition $\Phi_0(\cdot)=\Phi(0,\cdot)$ satisfying a growth assumption at infinity, e.g., at most linear growth.  If $\nu=0$, then the solution 
 $\Phi(t,x)$ is given by
the following formula:

\[
 \Phi(t,x)=\Phi^0(t,x)=\inf_{\gamma:\gamma(t)=x}\left[\Phi_0(\gamma(0))+\int_0^t L_\omega(\dot \gamma(s), \gamma(s),s)ds\right],
\]
where the infimum is taken over all absolutely continuous paths $\gamma:[0,t]\to\R^d$ satisfying $\gamma(t)=x$. 

In the viscous case $\nu>0$, this variational formula is replaced by a stochastic optimal control representation:
%\begin{equation}
% \label{eq:stoch-control-problem}
% \Phi(t,x)=\inf_{u} \E \left[\Phi_0(\gamma_t)+\int_0^t L_\omega (u(t-s,\gamma_s), \gamma_s, t-s)ds   \right],
%\end{equation}
%where $u(\cdot,\cdot)$ is a Markov control and $\gamma$ is the solution of the following SDE:
%\begin{align}
%\label{eq:controlled-sde}
% d\gamma_s&=u(t-s,\gamma_s)ds+\sqrt{2\visc}\,dB_s,\\
% \label{eq:controlled-sde-initial}
%\gamma_0&=x.
% \end{align}
 
\begin{equation}
 \label{eq:stoch-control-problem}
 \Phi(t,x)=\Phi^{\nu}(t,x)=\inf_{u} \E \left[\Phi_0(\gamma_t)+\int_0^t L_\omega (u(s,\gamma_s), \gamma_s, s)ds   \right],
\end{equation}
where $u(\cdot,\cdot)$ is a Markov control and $\gamma$ is the solution of the following  SDE 
\begin{align}
\label{eq:controlled-sde}
 d\gamma_s&=u(s,\gamma_s)ds+\sqrt{2\visc}\,dB_s,\\
 \label{eq:controlled-sde-initial}
\gamma_t&=x,
 \end{align}
with respect to the natural filtration that increases in the reverse time.
 
 We stress that the standard Brownian motion $B$ is not related to the disorder in the potential $F$, and is used solely as an auxiliary object in this representation.  It is intuitive that in the limit $\visc \to 0$, two variational formulas
match. It can be shown that the stochastic optimal control $u$ is equal to the velocity field corresponding to $\Phi(t,x)$, see~\cite{Gomes-Iturriaga-Khanin:MR2241814}, \cite{Fleming-Soner:MR2179357}. 

Namely, $u$ can also be represented as the Legendre conjugate
\[
 u(t,x)=\nabla_p H(p,x,t)\Bigr|_{p=p(t,x)},
\]
of
\[
p(t,x)=\nabla\Phi(t,x).
\]
 Thus, in addition to its variational meaning, 
representation~\eqref{eq:stoch-control-problem}--\eqref{eq:controlled-sde-initial} can be understood as a self-consistency
condition.

The most studied example corresponds to quadratic Hamiltonian $H(p)=|p|^2/2$. In this case, $L(v)=|v|^2/2$,
and the momentum and velocity are equal to each other. The  Hamilton--Jacobi  equation is 
then
equivalent to the Burgers equation
\begin{equation}
\label{eq:Burgers}
 \partial_t u + u\cdot \nabla u=\visc\Delta u + f
 \end{equation}
on the velocity field $u=\nabla\Phi$, where $f=-\nabla F$. Here we use the traditional notation $u$ for velocity.

The quadratic case is special in many ways. In particular, if $H(p)=p^2/2$, then
the Hopf--Cole transformation 
\begin{equation}
\label{eq:Hopf-Cole-0}
\Phi(t,x)=- 2 \visc \ln Z(t,x) 
\end{equation}
reduces equation~\eqref{eq:general-Hamilton--Jacobi} to the following linear heat equation:
\begin{equation}
\label{eq:parabolic-model}
\partial_t Z(t,x)=\visc \Delta Z(t,x)+\frac{F(t,x)}{2\visc}Z(t,x),  
\end{equation}
that is often called the parabolic Anderson model and can be solved using the Feynman--Kac formula:
\begin{align}
 \label{eq:Feynman-Kac}
  Z[Z_0](t,x)&=\E e^{\frac{1}{2\visc}\int_0^{t}F(t-s,x+\sqrt{2\visc}B_s)ds}Z_0(x+\sqrt{2\visc} B_t)\\
\notag           &=\E e^{\frac{1}{2\visc}\left[\int_0^{t}F(t-s,x+\sqrt{2\visc}B_s)ds-\Phi_0(x+\sqrt{2\visc} B_t)\right]},
\end{align}
where $Z_0(\cdot)=Z(0,\cdot)$ and $\Phi_0(\cdot)=\Phi(0,\cdot)$ are the initial conditions and $B$ is a standard Brownian motion in $\R^d$. 

It is natural to interpret the Feynman--Kac formula as averaging with respect to the Gibbs state with
free measure given by the Wiener measure on continuous paths and Boltzmann weights obtained from the potential
energy of the path. Namely, we can
introduce the distribution $\Pp_{t,x}$ on continuous paths on $[0,t]$ absolutely continuous with respect to the Wiener measure in reverse time emitted at point $x$ at time $t$ with variance $2\visc$, with
density
\[
p_{t,x}^{\omega}(\gamma)=\frac{1}{Z[\ONE](t,x)}e^{\frac{1}{2\visc}\int_0^{t}F(t-s,\gamma_s)ds},
\]
where $\ONE$ is the initial condition identically equal to~$1$.
Then one can rewrite~\eqref{eq:Feynman-Kac} as averaging with respect to the polymer measure $P_{t,x}$:
\begin{equation}
 \label{eq:FK-via-polymers}
 Z(t,x)=\langle Z_0(\gamma_0)\rangle_{\Pp_{t,x}}Z[\ONE](t,x). 
\end{equation}

In case $d=1$ and the potential $F$ being space-time white noise, 
equation~\eqref{eq:general-Hamilton--Jacobi} with quadratic Hamiltonian is known as
the KPZ equation. In this case making sense of solutions of this equation that have to have very low regularity is a nontrivial problem related to recent work of Martin Hairer, see, e.g.~\cite{Hairer:MR3071506}.  By assuming that our potential is smooth in~$x$,
we avoid this problem altogether. This is a natural assumption since the universality phenomena related to the KPZ equation concern the large scale picture and microscopic details of the setup should not be essential.

In the definition of polymer measures as Gibbs distributions, the role of temperature is played by the viscosity parameter $\nu$. In particular, the zero-viscosity
limit for Hamilton--Jacobi equations can be understood via zero-temperature limits for the associated
polymer measures. This connection also extends to generalized polymer measures arising in the stochastic
control construction of solutions of general Hamilton--Jacobi equations discussed above.

We now turn our attention to the problem of global solutions. 
We are interested in the behaviour of solutions
over long time intervals. In order to see statistically stationary behaviour, we must consider the dynamics
on equivalence classes, so that at any given time two functions that differ
by a constant are considered identical.

There are two ways to look at this problem. The first one is more traditional. Due to 
the nature of white noise, one can
view the solution as a Markov process in appropriate functional space $\Lc$ and study the long-term statistical properties
by looking at stationary distributions of this Markov process. Another way is to describe the evolution by a 
skew-product structure on the space $\Omega\times\Lc$. One can define a nonrandom flow on this space by
\[
 \Theta^t(\omega,\Phi)=(\theta^t\omega, S^t_\omega\Phi),
\]
where $S^t_\omega$ is the solution operator for the  Hamilton--Jacobi  equation from $0$ to $t$ defined by variational formulas.
One can study
invariant measures of $\Theta^t$ with fixed marginal distribution $\Pp$ on $\Omega$. Any such measure 
$\mu$
can be represented as
\begin{equation}
\label{eq:disintegration-into-sample-measures}
 \chi(d\omega,d\Phi)=\Pp(d\omega)\mu_\omega(d\Phi).
\end{equation}

The measure $\chi$ is called physical and measures $\mu_\omega$ are called sample measures 
if the dependence of $\mu_\omega$ on $\omega$ is measurable with respect to 
$\Fc^{-\infty,0}$ which is generated by $(F(t,x))_{t\le 0, x\in\R}$, i.e., it is determined by the history of the forcing  
from $-\infty$ to the present $(t=0)$, and 
$\mu_{\theta^t\omega}= S^t_\omega\mu_{\omega}$.

Ledrappier--Young \cite{Ledrappier-LSY:MR968818} proved that that there is a natural one-to-one correspondence between invariant distributions for Markov
processes and physical invariant measures for the skew product. Namely, every invariant measure $\mu$ of the Markov
process can be represented as the $\Lc$-marginal of a measure $\chi$ from \eqref{eq:disintegration-into-sample-measures} and thus admits a representation
\[
 \mu(\cdot)=\int_{\Omega} \Pp(d\omega)\mu_\omega(\cdot).
\]

What is described above is a general setting for random dynamical systems. In the case of random  Hamilton--Jacobi  equation, the situation
is more special. It turns out that in this case, the conditional distributions $\mu_\omega$ are delta-measures
concentrated on particular global solution $\Phi_\omega(t,x)$ defined almost surely. Obviusly
the uniqueness of global solution can be valid only up to an additive constant. It also depends on an additional 
parameter~$b\in\R^d$ that can be viewed as the average momentum. Namely, under mild conditions on
the stationary potential $F(t,x)$,
the class of function of the form
$\phi(x)=b\cdot x+\psi(x)$ where $\psi(x)$ has sublinear growth as $|x|\to\infty$ is invariant under the  Hamilton--Jacobi  dynamics.
Although we consider global solutions without initial conditions, they still ``remember'' the parameter~$b$. In other
words for every $b\in\R^d$, there is a unique global solution $\Phi_{b,\omega}(t,x)=b\cdot x+\psi_{b,\omega}(t,x)$,
where $\psi_{b,\omega}$ is sublinear in $x$.

The property where a unique point $\Phi_{b,\omega}(t,\cdot)$ is compatible with the history of the forcing is called
 One Force --- One Solution Principle (1F1S).
In other words, for fixed $b$, the global solution $\Phi_{b,\omega}(t,\cdot)$ is a deterministic
functional of the forcing up to time~$t$.
  This property immediately implies uniqueness of invariant distribution $\chi$
 for the skew-product system  and hence uniqueness of a 
 stationary distribution $\mu$ for the associated Markov process.

The real meaning of 1F1S is that the global solutions play the role of one-point random pullback attractors. Namely, solutions of the  Hamilton--Jacobi equation with initial data of the form
$b\cdot x+\psi(x)$
given at a time $-T$ converge to the global solution $\Phi_{b,\omega}(t,\cdot)$ as $T\to\infty$.

1F1S has been established for random dynamics defined by equation~\eqref{eq:general-Hamilton--Jacobi} in several settings. 

In~\cite{ekms:MR1779561},
1F1S was obtained in the inviscid setting for quadratic Hamiltonian and space-periodic forcing 
potential under some nondegenracy conditions. 
These results were clarified and expanded to the multidimensional periodic situation and other 
convex Hamiltonians in~\cite{Iturriaga:MR1952472}, \cite{Gomes-Iturriaga-Khanin:MR2241814}, \cite{Boritchev-Khanin}, \cite{KeZhang-Khanin2017}. In all these cases, the function $\psi_{b,\omega}$ introduced above is periodic in space.

Convergence of positive viscosity invariant measures to zero viscosity invariant measure for the torus case was also obtained in~\cite{Gomes-Iturriaga-Khanin:MR2241814}.  In these situations,  the problem is  effectively compactified since
one works on the circle or torus. Another similar result in compact setting is 1F1S  
for Burgers equation with random boundary conditions~\cite{yb:MR2299503}.

In the noncompact case, 1F1S has been established in 1D for quadratic Hamiltonian in the viscous and inviscid situations under certain assumptions on the forcing~\cite{Bakhtin-quasicompact}, \cite{BCK:MR3110798}, \cite{kickb:bakhtin2016}, \cite{Bakhtin-Li:2016arXiv160704864B},
\cite{Bakhtin-Li:preprint}.  It also can be proved in higher-dimensional  ($d\ge 3$) quadratic Hamiltonian case with positive viscosity and weak forcing. This special situation is called weak disorder and was studied starting with 1980's \cite{Imbrie-Spencer:MR968950},
\cite{Bolthausen:MR1006293}, \cite{Sinai:MR1341729}, \cite{Kifer:MR1452549}, \cite{Comets-Shiga-Yoshida:MR1996276}.

One of the main conjectures we want to formulate in this paper is that 1F1S is valid in full generality: 

\begin{conjecture}\label{conj1} For any $d\in\N$,  and any  $F_\omega(t,x)$ with exponential decay of space-time correllations, 1F1S
holds, i.e., for every $b\in\R^d$, there is a unique  time-stationary (modulo time-dependent additive constants) global solution $\Phi_{b,\omega}(t,x)=b\cdot x+\psi_{b,\omega}(t,x)$ where $\psi_{b,\omega}$ has sublinear growth. These solutions are continuous in $\nu\ge0$. In particular, global solutions 
are preserved under the zero-viscosity limit.
\end{conjecture}

Due to randomness, the role of $b$ is not essential in this picture in contrast with the Aubry--Mather theory where
arithmetic properties of $b$ play a crucial role. Often, if the value of $b$ is not indicated, we 
assume $b=0$.

The uniqueness of global solution of  the Hamilton--Jacobi equation is tightly related to the problem of existence and uniqueness of one-sided minimisers or geodesics in the inviscid case and one-sided directed polymers in the viscous case. 

Below we discuss this connection in the inviscid case. The Lax--Oleinik variational principle deals with minimisers defined on finite time intervals. For global solutions, one has to consider minimisers on intervals of the form $(-\infty,t]$. A curve is called a one-sided minimiser if any compact perturbation of the curve that does not affect the endpoint and the infinite tail,
can only increase the action. The counterpart of the conjecture above is the following conjecture.

\begin{conjecture} Under the same assumptions as in Conjecture~\ref{conj1}, almost surely the following holds:
\begin{enumerate}[(1)]
\item
Every one-sided minimiser $\gamma$ has an asymptotic slope
\[
 a=\lim_{s\to-\infty}\frac{\gamma(s)}{s}\in\R^d.
\]
\item For fixed $a\in\R^d$ and all $(t,x)$, there is a minimiser with endpoint $(t,x)$ and asymptotic slope~$a$. 
\item 
For fixed $a\in\R^d$ and Lebesgue-a.e.~$(t,x)$ a minimiser is unique.
\item For every two minimisers $\gamma_1,\gamma_2$ with the same slope,
\[
 \lim_{s\to-\infty}|\gamma_1(s)-\gamma_2(s)|=0.
\]

\item Let us fix $a\in\R^d$ and define
\[
\Phi^a_\omega(t,x)=\int_{-\infty}^t L_\omega(\dot\gamma_{t,x}(s),\gamma_{t,x}(s),s)ds-
\int_{-\infty}^0 L_\omega(\dot\gamma_{0,0}(s),\gamma_{0,0}(s),s)ds,
\]
where $\gamma_{t,x}$ is a minimiser with slope $a$ and endpoint $(t,x)$. 
Then $\Phi^a$ is a global solution of  Hamilton--Jacobi  and there is a uniquely defined $b=b(a)\in\R^d$ such that
$\Phi^a=\Phi_b$.
\item The function~$\Phi^a$ defined above is continuous and locally Lipschitz. Points $(t,x)$ of non-uniqueness
of one-sided minimisers are called shocks. They are discontinuity points of $\nabla\Phi^a$.  
\end{enumerate}
\end{conjecture}

The solution $\Phi=\Phi^a$ is stationary in time if we consider it modulo additive constants.  For  fixed~$t$, the field $\Phi(t,x)$ is not stationary in space but it has stationary spatial increments. In 1D, this allows to formulate the
following conjecture:

\begin{conjecture}\label{eq:CLT-conjecture}
Assume that $a=b=0$. Then there is a constant $\sigma>0$ such that the rescaled process $(\Phi(0,sx)/(\sigma\sqrt{s}))_{x\in\R}$ converges in distribution to the standard two-sided Wiener process as $s\to+\infty$.
\end{conjecture}

If we assume that the Hamiltonian $H$ is an even function of $p$, then, by symmetry $b(0)=0$. In general, the connection
between $b$ and $a$ is determined by the effective Hamiltonian which is the Legendre transform of the shape function. We will discuss the shape function and its connection to the
effective Hamiltonian later. Notice that in the general case, one should subtract the linear form $bx$ in the above conjecture.

We finish this section with an argument explaining why minimisers with different endpoints must be asymptotic to each other in the reverse time. The argument can be applied in any dimension.

Assume again that $a=0$ and consider ``point-to-line'' minimisers on a finite time interval $[-T,0]$, where $T\gg 1$.
Consider points $x$ and $y$ at distance of order $1$ apart. With large probability the difference
in action of minimisers with enpoints $(0,x)$ and $(0,y)$ is of order 1, since otherwise the
action of one of them can be decreased by merging with the other one.
On the other hand the variance of action of each minimiser grows with $T$. Hence, the difference 
of actions is of order $1$ only if these minimisers coincide or get close to each other for large negative times.

\section{Generalized directed polymers}
\label{sec:HJ-polymers}

In the previous section, we defined directed polymers associated with the  Hamilton--Jacobi  equation with quadratic
Hamiltonian and positive viscosity.  

Recall that the polymers were defined as paths on interval $[0,T]$.
Similarly to the situation with one-sided minimisers, it will be convenient to consider polymers to evolve backwards in time on intervals
$[-T,0]$.

The main goal of this section is to extend this notion
to more general Hamiltonians where the Hopf--Cole transformation is not available. Asymptotic properties of the classical directed polymers were studied in  \cite{Carmona-Hu-2004:MR2073415} and \cite{Comets-Shiga-Yoshida:MR1996276}.
Their findings
can be summarized in the following statement. 

{\bf Proposition.} {\it In dimensions $d=1$ and $d=2$, the directed polymer is localized. In dimension $d\ge 3$, there is a transition between
diffusive behavior for small forcing potentials and localization for large forcing.  
}

Rigorous mathematical methods allow to prove only weak forms of localization, namely, that there is a constant $\alpha>0$ not depending on time and a family of arbitrarily large times $t$ and  intervals~$I_\omega(t)$ of length~$1$ such that the probability to find the endpoint of the polymer in $I_\omega(t)$ is at least $\alpha$. However, the localization picture
to have in mind is that for typical $\omega$ the variance of the endpoint
of the polymer with respect to the polymer measure is bounded. A subsequential version of this claim was recently established in~\cite{Bates-Chatterjee:2016arXiv161203443B}.
This is in
sharp contrast with diffusive behavior where the variance of the endpoint
is of order~$T$.

Let us now define generalized polymers for general Hamiltonians. For a fixed~$\omega$ at $T>0$, the directed polymer on time interval $[-T,0]$ is
the stochastic process $X$ that solves the following SDE in reverse time
\begin{align}
\label{eq:gen-poly}
dX_s&=u(s,X_s)ds+\sqrt{2\nu}dB_s,\\ 
X_0&=x,\notag
\end{align}
where $u(\cdot,\cdot)$
is the velocity field corresponding to the solution $\Phi$ of the  Hamilton--Jacobi  equation on $[-T,0]$ with $\Phi(-T,x)\equiv 0$:
\[
 u(s,x)=\nabla_p H(\nabla\Phi(s,x)). 
\]
We call the distribution of $X$ the generalized polymer measure. 
Naturally, one can consider polymers associated to more general initial conditions $\Phi(-T,\cdot)$.

In the quadratic Hamiltonian case, the generalized polymer measures
coincide with the usual Burgers--Hopf--Cole--Feynman--Kac--Gibbs polymer
measures. This
is an important fact that justifies the new notion. 
We postpone a detailed explanation
of this connection to
Section~\ref{sec:equiv-of-poly-for-Burgers}.

It is also possible to state the positive viscosity versions of conjectures in 
Section~\ref{sec:HJ-eq} in terms of generalized polymers. The role of one-sided minimisers in this
positive temperature situtation is played by one-sided polymer measures that can be viewed as thermodynamic limits of finite volume directed polymers. These notions are classical in the case of quadratic Hamiltonian, see~\cite{Bakhtin-Li:2016arXiv160704864B}, and 
can be naturally extended to general Hamiltonians and associated polymers.  Moreover, we conjecture that
in the zero-temperature limit $\nu\to 0$, the generalized polymers converge to their respective ground states, i.e., one-sided Lagrangian minimisers, see~\cite{Bakhtin-Li:preprint} for the quadratic Hamiltonian case.

Also based on the analogy with the quadratic case, it is natural to state the following conjecture:

\begin{conjecture} In dimensions $d=1$ and $d=2$, the generalized directed polymer is localized for any convex Hamiltonian with superlinear growth of the associated Lagrangian. In dimension $d\ge 3$, the generalized polymer undergoes
a transition from
diffusive behavior for small forcing potentials to localization for large forcing.  
\end{conjecture}

Although the description of the generalized polymers through SDEs is less explicit than the one provided by the Feynman--Kac formula, it has significant advantages as well. First, the drift $u$ in~\eqref{eq:gen-poly} is the optimal
control in the variational principle~\eqref{eq:stoch-control-problem}.
Secondly, this drift is given by the velocity field which has a clear physical meaning and provides an
intuitive explanation of the localization phenomenon.

It is well-known that in the inviscid case, the velocity
field pushes points towards shocks. When viscosity is positive, the velocity still pushes them towards compact shock-like regions working against the diffusion effect provided by white noise. If the shock-like domains are periodic in space (say, in the periodic case), the polymer would still have diffusuve behavior since transitions between those regions, although rare, happen with positive rate, so over large times the process can be viewed as a random walk over those regions. However, if, due to fluctuations, the polymer reaches a domain with isolated shock-like region evolving in time, then the velocity field will provide a potential barrier that the polymer will not be able to escape through. Hence the polymer will be localized near a time-dependent zone concentrated around the shock.
So, in contrast to the periodic case where the transition rates between neighboring shock-like regions are constant which leads to diffusive behavior, in the nonperiodic case the
typical escape times vary from trap to trap thus resulting in 
an situation similar to Sinai's random walk in random environment where deep traps create a localization effect.

In dimensions starting with~$3$, however, there is also another effect that
has to be taken in consideration. A small potential barrier cannot trap
diffusing particles. Hence for small velocity fields generated by small potentials the polymer will have diffusive behavior. This explains the
transition from weak disorder to strong disorder in high dimensional case.

\section{Shape function and KPZ scalings}
\label{sec:shape}
In this section, we discuss the shape function which plays an important role in all the discussions below. To define it one has to consider either point-to point minimisers  or directed polymers
conditioned to end at a certain point. In other words, consider a minimiser $\gamma$ on a
time interval $[-T,0]$ with boundary conditions $\gamma(0)=0, \, \gamma(-T)=-aT$, or a directed polymer 
corresponding to $\nu>0$
with the same boundary conditions. We start with the case of quadratic Hamiltonian.
We denote the Lagrangian action of the minimiser by $A_\omega(a, T)$ and the corresponding partition function
by 
\[
Z_{\nu,\omega}(a,T)=\int e^{\frac{1}{2\nu}\int_{-T}^0 F_{\omega}(s,B_s)ds}P_{\nu,aT,0}^{-T,0}(dB),
\] 
where $P_{\nu,-aT,0}^{-T,0}$ is the distribution of the Brownian bridge of variance $2\nu$ connecting points $-aT$ and $0$ between times $-T$ and $0$.
Then the following general statement follows from the sub-additive ergodic theorem.

\begin{statement*}  For any $\nu\ge 0$, there is a
nonrandom convex function $S_\nu(a)$ such that for every $a\in\R^d$, 
almost surely
\begin{align*}
&\lim_{T\to \infty} \frac{A_\omega(a, T)}{T} = S_0(a),\\
&\lim_{T\to \infty}\frac{-2\nu\log Z_{\nu,\omega}(a, T)}{ T} = S_\nu(a),\quad \nu>0.
\end{align*}
\end{statement*}

The problem with the above statement is that it is very difficult to control the shape functions
$S_\nu(a)$. In most applications, one needs to know that they are strictly convex and have a non-degenerate minimum at the origin. Although there are no doubts that this
should be true in a very general situation, including general Hamiltonians, at present we don't have proper technical tools to
prove such a statement. In all the cases where this can be done, one uses special symmetries which
allow to get an exact formula for  $S_\nu(a)$.  
One such special property 
is the  so called shear-invariance. The random potential $F_\omega(t,x)$ 
is said to be shear-invariant if for all $a \in \R^d$, the  processes
$F_\omega(t,x+at)$ and~$F_\omega(t,x)$ have the same distribution. This property is
satisfied for many natural models. In the case of quadratic Hamiltonian one can calculate $S_\nu(a)$ explicitly provided the shear-invariance holds. It turns out that
$$ S_\nu(a)= S_\nu(0) + \frac{|a|^2}{2},\quad \nu\ge 0,\ a\in\R^d.$$
Below we deal with the case of minimisers. The polymer case is similar. Notice that any curve $\gamma_a$ going from
the origin to the point $-aT$ on the time interval $[-T,0]$ can be obtained from a curve $\gamma$ starting and
terminating at the origin, by adding a linear function $at$. By the shear invariance the distribution for the action of $\gamma$ and $\gamma_a$ is the same apart from the 
kinetic term. Since
$$\int_{-T}^0\frac{|\dot \gamma_a|^2(s)}{2}ds= \int_{-T}^0\frac{|\dot \gamma (s)+a|^2}{2}ds=
\int_{-T}^0\frac{|\dot \gamma(s)|^2}{2}ds + \frac{a^2}{2}T,$$
it follows that $S_0(a)= S_0(0) + |a|^2/2$.

For general Hamiltonians, the definition of shape function for positive viscosity must be modified and can be based
on the linear growth of solutions of the associated Hamilton--Jacobi equation with initial condition concentrated near $-aT$ at time $-T$.  Even for shear-invariant potentials strict convexity
of the shape function is an open question.

Notice that the shape functions are closely related to the concept of homogenisation for random
Hamilton--Jacobi equations. Since the slope $a$ equals the average velocity, the shape function
can be considered as the effective Lagrangian. Then its Legendre transform is the
effective Hamiltonian~$H_{\eff}$.
To clarify this connection, consider the inviscid Hamilton--Jacobi equation with initial
condition given at $-T$ by $b\cdot x+\psi(x)$ for a 
 a fixed $b\in \R^d$.
To find the slope $a\in \R^d$ corresponding to~$b$, notice that for a given slope $a$ the
action on the time interval $[-T,0]$ will be $(S_0(a) - a\cdot b)T$ plus terms sublinear in~$T$. 
Minimizing $\min_a[S_0(a) - a\cdot b]$ we obtain $a(b) = \argmax (a\cdot b - S_{0}(a) )$ and $H_{\eff}(b)= \max_a[a\cdot b - S_{0}(a)]$. This also implies that $a(b)=\nabla H_{\eff}(b)$.

\begin{figure}
  \includegraphics[width=8cm]{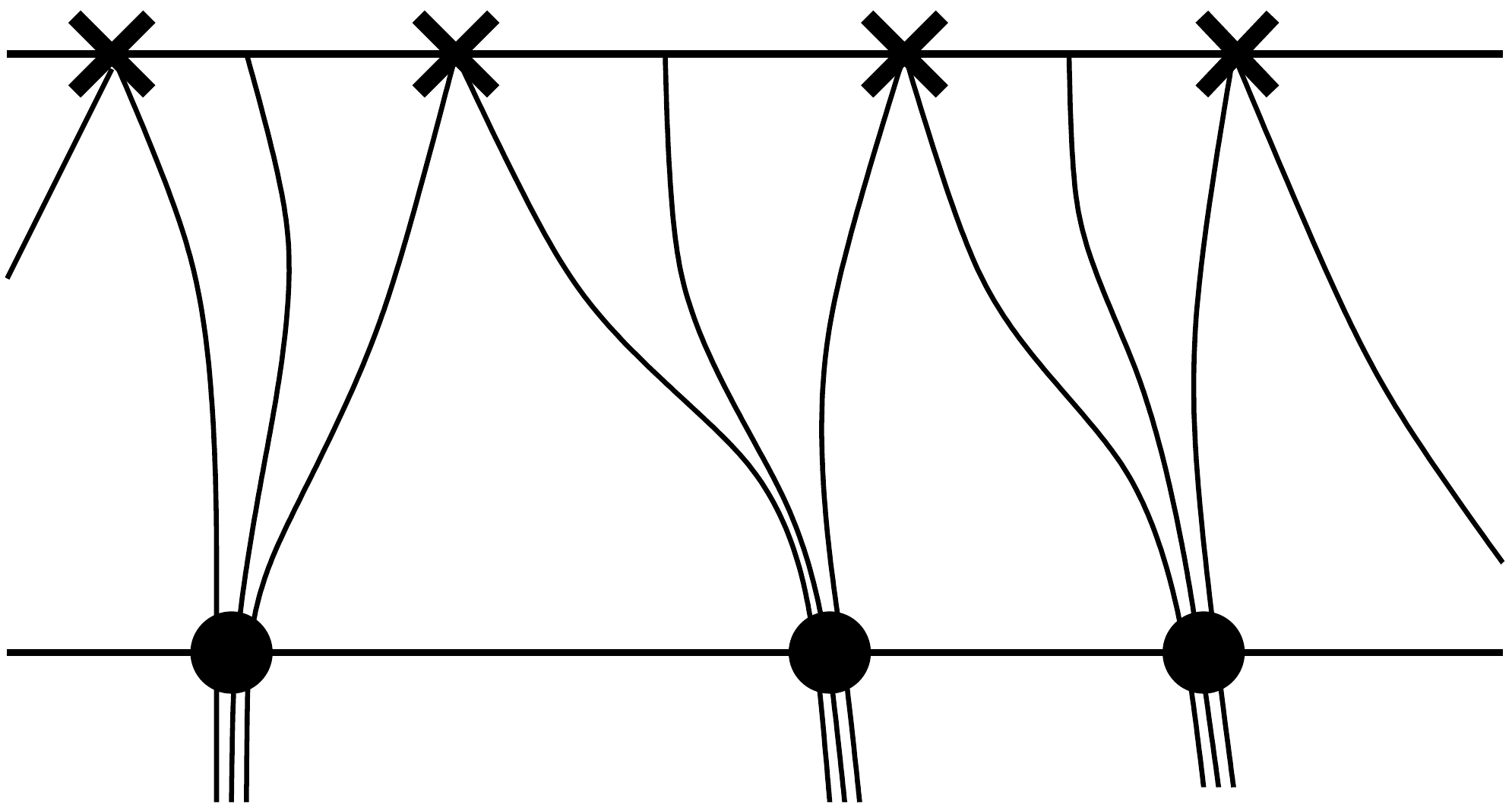} 
\caption{Long-time behaviour of one-sided minimisers. The time axis is directed upward. The dots repesent locations of high concentration of minimisers. The crosses represent shocks separating the minimisers.}
\label{fig:shock-minimiser-duality}
\end{figure}

We now switch to the one-dimensional case and demonstrate how the KPZ scaling exponents
follow from our conjectures and the quadratic behaviour of the shape function. What makes the 
one-dimensional case so special is the property of monotonicity. Minimisers cannot cross each other
and maintain the same order at all times.
%\sout{, so at all times a one-sided minimiser which has the end-point to the right of the end-point of another one will
%stay to the right of it backward in time.} 
On the other hand, all one-sided minimisers are asymptotic
to each other  backwards in time. This suggests the following picture. Consider $T \gg 1$ and
all one-sided minimisers on time interval  $(-\infty, 0]$. By time $-T$ the minimisers will concentrate 
in small (exponential in $T$) intervals separated by large intervals. Each small interval of concentration
generates a large interval of minimiser endpoints at time $t=0$. We will discuss this picture in more details in the next section. Considering the small concentration intervals as points we get two random point fields,
one at time $t=0$,  and another at time $t=-T$. The first point field consists of points of separation
between long intervals corresponding to neighbouring concentration domains.  Those separation points are, in fact, points of shocks, since each of this points has at least two minimisers  going
to different concentration domains (see Figure~\ref{fig:shock-minimiser-duality}).

Denote by $L(T)$ a characteristic length corresponding to time scale $T$. It can be defined as a typical distance between neighbouring 
points of any of the two point fields described above. It can also be viewed as a typical displacement
of the one-sided minimiser on time interval of length $T$. Denote by $\xi$ the corresponding critical
exponent, namely $L=L(T)\sim T^\xi$. The second critical exponent $\chi$ corresponds to typical
fluctuations of the action of one-sided minimisers on time intervals of length~$T$ which are
assumed to be of the order of $T^\chi$. We shall show below that $\xi=2/3$, $\chi=1/3$.
Sometimes these scalings are referred to as 1:2:3 scalings. 

Let us compare the one-sided minimiser with endpoint $(0,0)$ to the optimal path~$\bar\gamma$ on $[-T,0]$ beginning and terminating at the origin. If the displacement of the one-sided minimiser is of the order of $L$, then, compared to $\bar\gamma$, its action increases by a term of the order of~$L^2/T$ due to the quadratic behaviour of the shape function, but it can also decrease due to the fluctuations of the
action on time interval $[-T,0]$ and fluctuations of the tail action on the interval $(-\infty, -T]$.
The first of these fluctuations is of the order~$T^\chi$, while the second is of the order
of $\sqrt{L}$ due to the conjectured diffusive scaling for the stationary solutions. The scale $L$
is determined by either assuming that 
\begin{equation}
\label{eq:first-term-dominates}
L^2/T \sim \sqrt{L}\quad \text{and}\quad T^\chi\lesssim \sqrt{L},
\end{equation} 
or, in the opposite case, 
\begin{equation}
\label{eq:second-term-dominates}
L^2/T \sim T^\chi,\quad \text{and}\quad \sqrt{L}\lesssim T^\chi.
\end{equation} 
In the case $\chi > 1/3$ the first assumption 
of \eqref{eq:first-term-dominates} leads to $L\sim T^{2/3}$ and, hence,
$T^\chi \gg \sqrt{L}$ which contradicts the second condition of~\eqref{eq:first-term-dominates}. 
It follows that the first assumption of~\eqref{eq:second-term-dominates} holds,
which implies $L\sim T^{(1+\chi)/2}$ and $\sqrt{L}\sim T^{(1+\chi)/4}\ll T^\chi$. This, however,
contradicts the predicted diffusive behaviour for the global solution. Indeed, suppose that points $x$ and $y$ are at the distance of the order of~$L$ apart, then, since the fluctuations of each $\Phi_\omega(0,x)$, $\Phi_\omega(0,y)$ are of the order of~$T^\chi$, 
the difference $\Phi_\omega(0,x) -\Phi_\omega(0,y)$ will be at least of order
$T^\chi$ which is much larger than $\sqrt{L}$. 
We conclude that $\chi\leq1/3$.  It is easy to see that in this case both
~\eqref{eq:first-term-dominates} and \eqref{eq:second-term-dominates} lead to $L^2/T \sim \sqrt{L}$, and, hence, $L\sim T^{2/3}$ and $\xi=2/3$. 

Let us  show that $\chi = 1/3$. Suppose that $\chi< 1/3$ and consider two points $x$ and $y$ at distance of the order of $L$ apart. On the time scale of the order of $T\sim L^{3/2}$ their minimisers will almost merge with positive probability. On a slightly larger time scale $T\sim L^{3/2+\epsilon}$, the minimisers will merge
with large probability. Then the difference $\Phi_\omega(0,x) -\Phi_\omega(0,y)$  will be
determined by two contributions: one from the shape function, and another from the
fluctuation of action on time interval $[-T,0]$. The position of the concentration point at time
$-T$ will be of the order of $T^{2/3}\sim L^{1+\frac{2}{3}\epsilon}$.  Hence the shape function contribution
will be of the order of $L\cdot L^{1+\frac{2}{3}\epsilon}/T\sim  L^{2+\frac{2}{3}\epsilon}/T\sim L^{\frac{1}{2}-\frac{1}{3}\epsilon}$. The second contribution
will be of the order of $T^\chi\sim L^{(3/2 +\epsilon)\chi}$. Now one can take $\epsilon$ so small
that the exponent satisfies $(3/2 +\epsilon)\chi< 1/2$. Thus $|\Phi_\omega(0,x) -\Phi_\omega(0,y)|\ll
\sqrt{L}$ which again contradicts the diffusive behaviour assumption.

\section{minimisers and shocks in the 1D case}
\label{sec:min-shocks-1d}
In this section, we consider in more detail the behaviour of one-sided minimisers and 
shocks in the one-dimensional case. In the previous section, we already mentioned 
the monotonicity property that means that minimisers do not intersect. Here we present
a more complete picture.

Let us consider the field of one-sided minimisers with fixed, say, zero slope.

Let us consider two times $t_1<t_2$ and denote space variables at these times by $y$ and $x$, respectively. If there is a unique one-sided minimiser with the end-point $(t_2,x)$ then there is a unique $y(x)$ such that this minimiser passes through $(t_1,y(x))$. On the other hand,
if $(t_2,x)$ is a point of shock so that there is more than one minimiser at this point,
then there is a closed interval $[y_l(x),y_r(x)]$ of points at time $t_1$ 
associated to $(t_2,x)$. Namely, these are points absorbed into the shock~$(t_2,x)$.
Here the point $(t_1,y_l(x))$ corresponds to the left-most minimiser with endpoint~$(t_2,x)$, while 
$(t_1,y_r(x))$ corresponds to the right-most minimiser.  The map from $x$ to $y$ is monotone, injective,
but, in general, not single-valued.  One can also construct the map from $y$ to $x$.
Namely, if a minimiser with endpoint $(t_2,x)$ passes through $(t_1,y)$ then $x(y)=x$.
Also, for any shock point $(t_2,x)$, all points $y$ from $(y_l(x),y_r(x))$  are mapped into~$x$.
This map is also monotone, single valued, but not injective. 

Denote by $A(t_1,t_2)$ the set of points 
$(t_1,y)$ reachable by minimisers with endpoints $(t_2,x)$, $x\in\R$. It is easy to see that the sets $A(t_1,t_2)$ 
are decreasing in $t_2$, i.e., $A(t_1,t'_2)\subset A(t_1,t_2)$ as long as 
$t'_2>t_2$. If there is $(t_1,y) \in \cap_{t_2>t_1} A(t_1,t_2)$, then there is a global minimiser
passing through $(t_1,y)$. That means that the one-sided minimiser for $(t_1,y)$ can be
extended indefinitely forward in time keeping the property of being a one-sided minimiser.

If $(t_1, y)$ is a point of shock, then no one-sided minimiser defined at $t_2>t_1$ can pass through it.
Therefore, this point is covered by a closed interval generated by some shock at time $t_2$.
In other words, for every $t_2>t_1$ there exists a shock at $(t_2, x(y))$. That means that there is a curve of shocks $\{(x(y),t_2), t_2\geq t_1\}$, where $x(y)=y$ for $t_2=t_1$. In other words, every
shock evolves forward in  time, while one-sided minimisers evolve backward in time.
One-sided minimisers are asymptotic to each other in backward time but do not intersect.
The curves of different shocks can merge. 

In a certain sense, minimisers and shocks are dual objects. However, there is also a significant difference. Namely, shocks
can be created at certain space-time points, called pre-shocks. Pre-shocks correspond to
singularities of the velocity field. The space derivative of the velocity field at the point of pre-shock
is equal to~$-\infty$. The shock curves originating at a preshock point evolve in time and merge
with other existing shock curves while new shocks are being formed. It follows from 
the results of \cite{BCK:MR3110798}, \cite{kickb:bakhtin2016} for the quadratic Hamiltonian case 
and is conjectured for the general Hamiltonian case that the processes of
eliminating shocks through their merging, and creating new shocks through the pre-shock 
phenomenon are balanced, so that they produce an equilibrium
corresponding to a stationary distribution of shocks.  At a given time, shocks form a stationary
point field. In addition there is another parameter attached to every shock, namely, its age,
indicating for how long a shock can be traced in the past until the original pre-shock event.
In a stationary regime, the age of a shock is just a random variable with 
distribution given by some density $q_s(a)da$, where~$a$ is the age value. Although the density $q_s(a)$
is not universal, the behaviour of its tails reflects the universal KPZ scalings. We have seen above
that one-sided minimisers after long time $T$ concentrate at certain random locations,
with a point field of shocks separating intervals of minimisers corresponding to a given location. 
These shocks are separated by intervals of the length of order $T^{2/3}$, hence the density of
the corresponding point field is of the order of $T^{-2/3}$. It is easy to see that this sparce
separating point field corresponds to shocks of age $a$ of the order $T$ and larger. This
implies that the density $q_s(a)$ must decay as $a^{-5/3}$. 

Similarly to the discussion
of separating shocks, intervals of concentration of minimisers
correspond to locations where minimisers can be extended in the future for time order $T$ or larger.
We shall explain below that in the general case global shocks do not exist. In other words, every one-sided minimiser will terminate being absorbed by some shock at some future time. Hence, one can define a random variable of life expectancy of one-sided minimisers. In the stationary regime this random variable has a density $q_m(a)da$. An argument similar to the previous one implies that the density
$q_m(a)$ also decays as $a^{-5/3}$ as $a \to \infty$.

There is one case when a rigorous mathematical analysis  of the behaviour of shocks and minimisers
was carried out. This is the so called compact case. In the simplest situation, the random potential
$F_\omega(t,x) = \sum_{i=1}^N F_i(x)\dot W_i(t)$, where $F_i(x), 1\leq i\leq N$, are non-random smooth
$1$-periodic functions, and $\dot W_i(t), 1\leq i\leq N$, are independent white noises. 
Periodicity condition implies that the configuration space for the problem is the unit circle.To avoid 
degenerate behaviour we also assume that the map $(F_1,\ldots,F_N)$ from $\T^1=\R^1/\Z^1$ to $\R^N$ is an embedding. Then the following theorem holds (\cite{ekms:MR1779561},\cite{Boritchev-Khanin}). 

\begin{theorem}
For every $a\in \R^1$, almost surely there exists a unique global minimiser
with asymptotic slope $a$. It is
a hyperbolic trajectory of the random Lagrangian flow. All other
one-sided minimisers with slope $a$ asymptotically approach the global one backward in time.
Moreover, the rate of convergence is exponential and is given by a non-random positive
Lyapunov exponent.
\end{theorem}

In fact, this theorem also holds in multiple dimensions under the periodicity assumption, but the mechanism of hyperbolicity is very different and the proof in that case requires additional ideas,
see \cite{KeZhang-Khanin2017}.

\begin{figure}
\includegraphics[width=8cm]{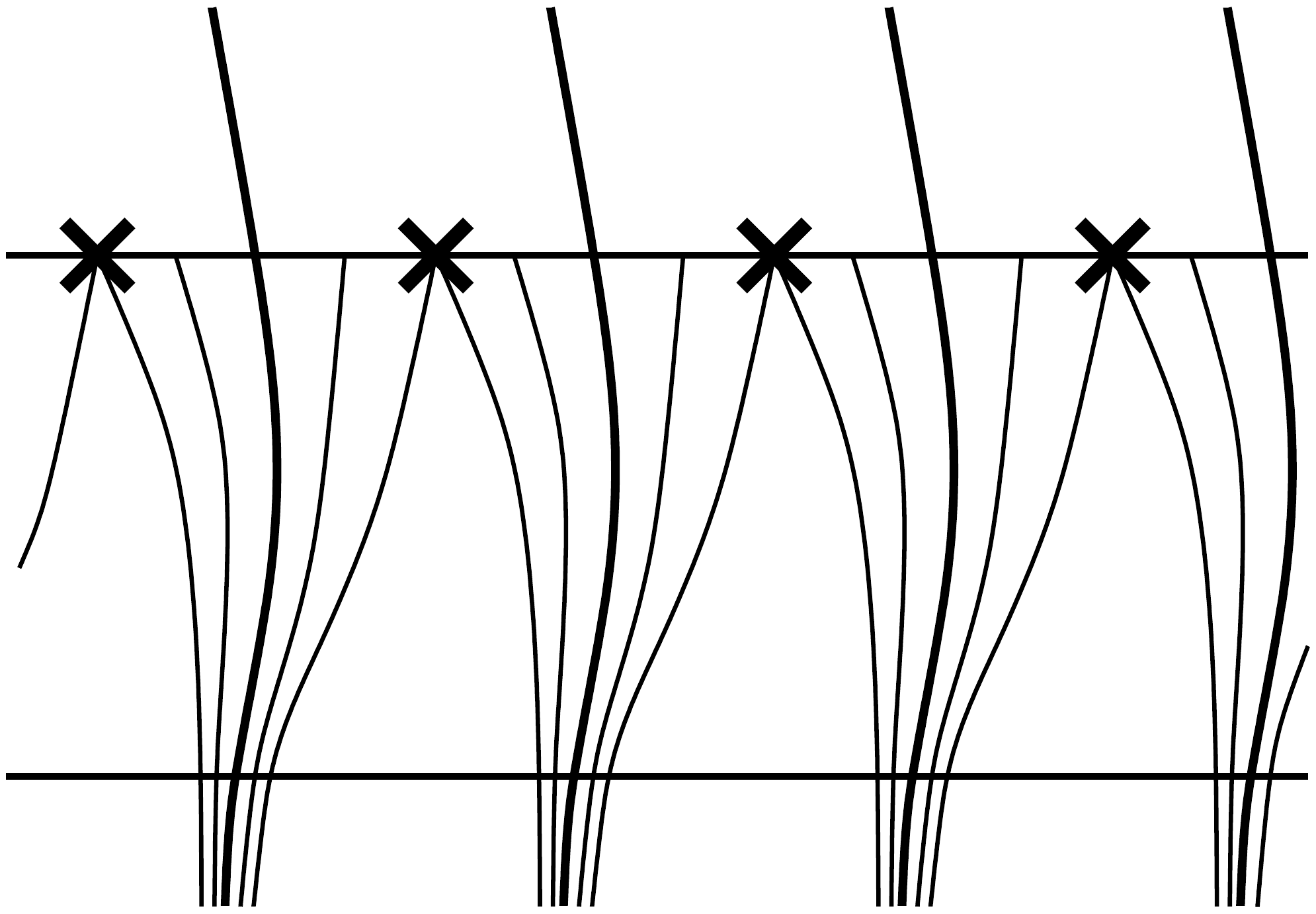} 
\caption{Minimisers and shocks on the universal cover in the periodic case.}
\label{fig:universal-cover}
\end{figure}
To better understand the structure of minimisers and shocks it is convenient  to lift the
picture to the universal cover $\R^1$ (see Figure~\ref{fig:universal-cover}). Different bold curves are different copies
of the global minimiser. All one-sided minimisers approach the global one but
they have to choose its particular representative to follow. Hence there are separation points that
correspond to the global (or, topological) shock. This shock can be traced indefinitely
in the past, and hence it is infinitely old. All other shocks are local in nature. This means
that although there are two minimisers meeting at a local shock, they are asymptotic to each other and
the same global minimiser in reverse time. 
All these shocks have finite age. However, the density $q(a)$ of their age decays exponentially fast 
as $a \to \infty$. This is in contrast with the behavior conjectured above for the nonperiodic case where $q(a)\sim a^{-5/3}$.  The unique global minimiser and the unique global shock exist due to
purely topological reasons. 

It is expected that in the general non-periodic case there are no
global minimisers and no infinitely old shocks. The reason for this is easy to see.
Recall that the sets $A(t_1,t_2)$ defined above shrink as $t_2$ increases.
The complementary set $\R^1 \setminus  A(t_1,t_2)$ of locations forbidden for one-sided minimisers
grows as $t_2 \to \infty$. It is natural to expect that any compact subset of $\R^1$
will belong to the forbidden set after some large time $t_2$. So, as $t_2$ increases there are simply
no place available for a global minimiser. 

\begin{figure}
\includegraphics[width=10cm]{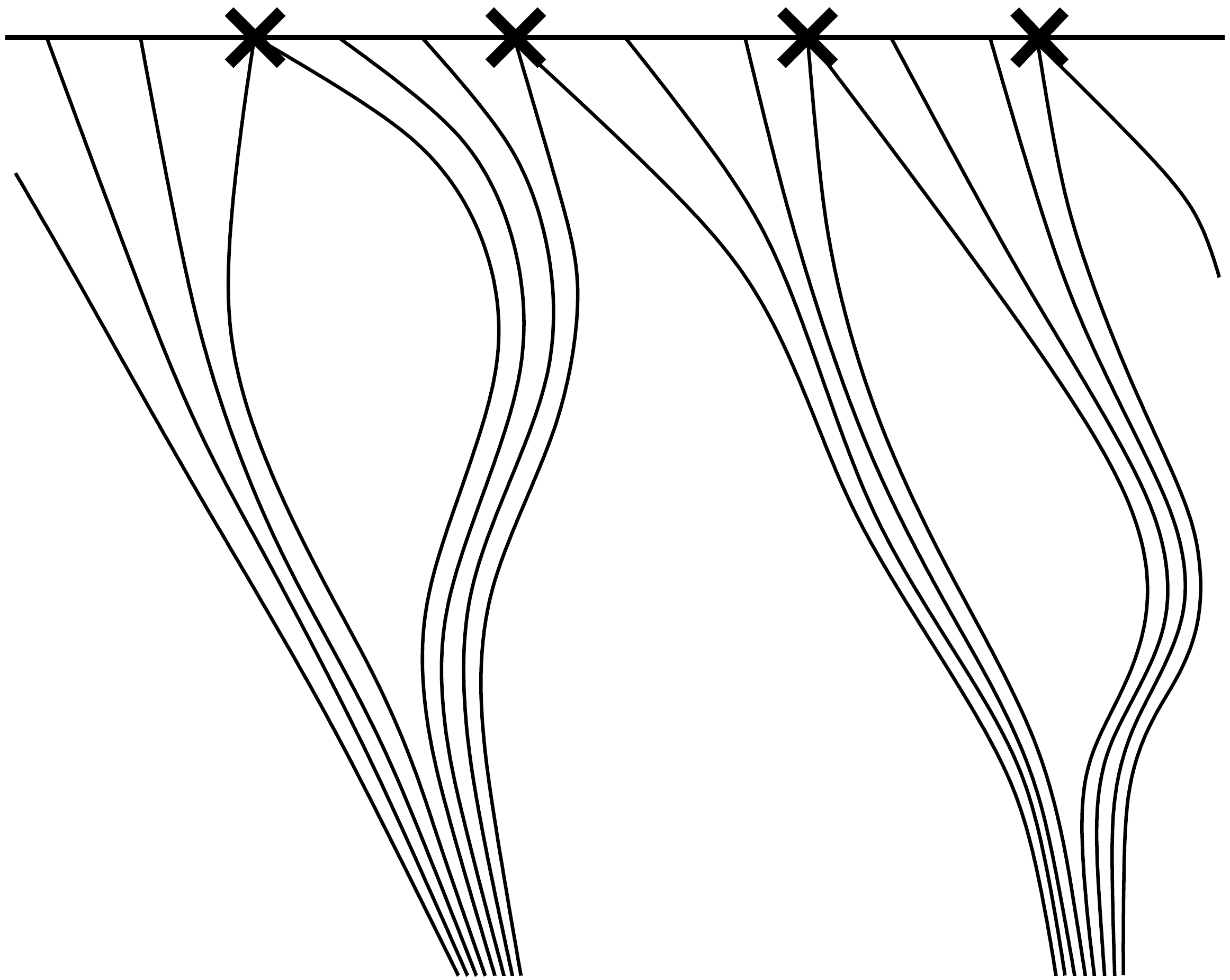} 
\caption{Minimisers and shocks in noncompact case}
\label{fig:clustering-of-minimisers}
\end{figure}

The conjectured behaviour of minimisers and shocks is somewhat similar to the periodic case, but has significant differences as well.
All one-sided minimisers are asymptotic to each other with the exponential rate of convergence.
However, two minimisers can be separated for a long time by an old shock. Eventually
they will come close and will start to converge exponentially, see Figure~\ref{fig:clustering-of-minimisers}. The time of separation $T$ and
the original distance between minimisers $L$ are connected by the KPZ scaling relation $T\sim L^{3/2}$. It is natural to expect that for large $L$ the distance between two
minimisers backward in time decays as
$\exp(-\lambda (|t| -T(x,y))_+ )$,
where $\lambda$ is a 
non-random Lyapunov exponent, and $T(x,y)\sim L^{3/2}$.

The picture described above is mostly conjectural. Only few results in this direction have been
proved for  few concrete models. We wanted to present it in order to emphasise two aspects that are not always discussed in the literature. The first one is the structural interaction between minimisers and shocks. The second one is the hyperbolicity of minimisers. Although the hyperbolicity is not
uniform, it is still a very essential feature of the problem. It is important that two different 
phenomena are present simultaneously: the separation by old shocks whose density
decays algebraically, and the exponential convergence when minimisers are not separated anymore.

\section {Renormalization for point fields of concentration and separation}
\label{sec:renorm-point}

In this section, we look more closely at the point fields of concentration and separation
defined above. 

\begin{figure}
\includegraphics[width=12cm]{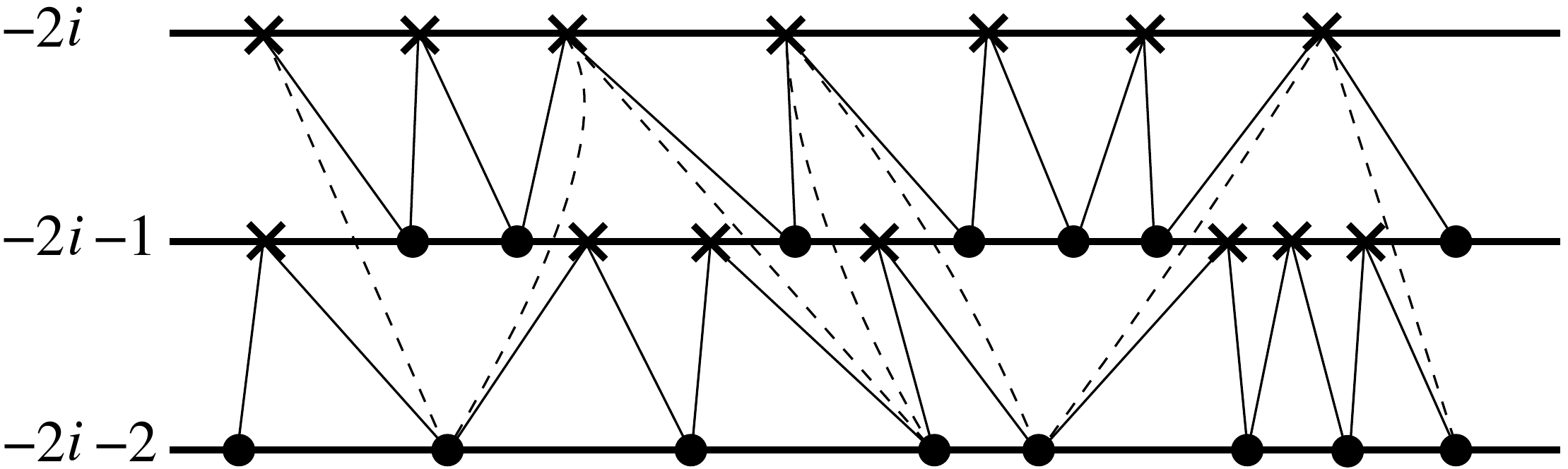} 
\caption{The renormalisation transformation on point fileds of crosses, dots, and connectors. The solid lines represent connectors before the transformation. The dashed lines represent the connectors
after the transformation. The crosses and dots with no adjacent dashed lines are eliminated under the renormalisation.}
\label{fig:points-and-crosses}
\end{figure}

Let us fix $T\gg 1$ and consider a sequence of times $-iT, \, i\geq0$.
For each time $-iT$, we consider all one-sided minimisers with endpoints $(x,-iT), x\in \R^1$.
As before, we record small intervals of concentrations of these minimisers at time $-(i+1)T$
and the separating point field at time $-iT$. As we saw above, these point fields will have density
of the order of $T^{-2/3}$.  Let us rescale them by $const\,T^{2/3}$ so that the resulting
point fields will have density~$1$. We also rescale time by a factor $T$ so that we have
two point fields at every strip $[-i, -i-1]$ (see Figure~\ref{fig:points-and-crosses} where shocks are denoted by
crosses, and concentration points by dots). 

Another important piece of this structure
is the connector field. We can enumerate intershock intervals by 
integers~$\Z$, from left to right, assigning $0$ to the interval containing the origin. Similarly, we can enumerate the concentration points assigning $0$ to the point closest to the origin. Every interval between two neighbouring shocks corresponds to a particular point of concentration, thus
defining a map $\Z\to\Z$. Since this map is monotone,
it is enough to determine the label of the point which corresponds to $0$-interval. 
We denote this integer number by $\xi$, and define a related sequence of random variables
$\xi_i$, called connectors, for every strip $[-i, -i-1]$. Notice that the sequences of
point fields and connectors for different strips form a stationary but strongly correlated
sequence. In addition, two point fields corresponding to any strip are also strongly correlated.

We next define a renormalisation transformation corresponding to rescaling the time by 
the factor of two. The transformation can be defined separately on individual 
strips of width 2 in time units of the previous step.
Namely, out of a configuration of crosses, dots, and connectors in each strip 
of the form $[-2i,-2i-2]\times\R, \, i\ge 0$, we will obtain a new configuration
of crosses, dots, and connectors in the strip $[-i,-i-1]\times\R$.

After the renormalisation transformation, 
a cross can be present at $(-i,x)$  only if there was a cross
at $(-2i,x)$ before the renormalisation. Some crosses survive the renormalisation,
and some get eliminated. Also, a dot can be present at
$(-i-1,x)$  only if there was a dot
at $(-2i-2,x)$ before the renormalisation. Some dots survive   the renormalisation,
and some get eliminated. The point fields of dots and crosses at odd times $-2i-1,$ $i\ge 0$, will be discarded but they are used to decide which crosses on $\{-2i\}\times\R$ and
dots on $\{-2i-2\}\times\R$ survive the renormalisation and which ones get eliminated. The
elimination procedure that mimics the behaviour of
shocks and minimisers is the core of the transformation. It can be described as follows.

We start with a point field of crosses at time $-2i$,  both point fields, of 
crosses and dots, at time $-2i-1$, and a point field of dots at time $-2i-2$. In addition,
each interval between two crosses at time $-2i$ is mapped to a particular dot at time $-2i-1$,
each interval between two crosses at time $-2i-1$ is mapped to a particular dot at time $-2i-2$.
Consider now an interval between two neighbouring crosses  at time $-2i-1$.

If there are no dots inside this interval, then the dot corresponding to this interval
at time $-2i-2$ should be eliminated. Indeed, no one-sided minimisers at time $-2i$ will
reach the interval between crosses, so the concentration point to which the
interval is mapped, also cannot be reached. 

Now, suppose that there are $k$ dots inside
the above interval between crosses. At time $-2i$, there are $k+1$ neighbouring
crosses corresponding to these $k$ dots. It is easy to see that for consistency with
behaviour of shocks and minimisers, we have to eliminate all
but the first and the last cross so that the whole interval gets mapped into one dot at time $-2i-2$. The other crosses do not separate anything anymore and have to be eliminated. Note
that if $k=1$, we will have only two crosses upstairs, the first and the last, and
we don't eliminate anything at all (see Figure~\ref{fig:points-and-crosses}). This elimination procedure should be carried out for all strips. After that, one should rescale time by the factor~$2$, and rescale point fields so that they will
again have density~$1$. Note that the new connectors are uniquely defined by the above procedure. The result is one step of the renormalisation transformation $R$. Note that
the rescaling can be done after several steps since it commutes with the renormalisation procedure.
Moreover, due to this commutativity, one can naturally define renormalisation in the space of point fields modulo rescaling. 

Since at 
each step we double the time interval, iterating the transformation we defined above results in exponentially growing time intervals.  One can
also define it for linearly growing time intervals. In this case one should add strips one by one.

Suppose we applied the renormalisation by adding together $n$ strips of the initial width one.
The resulting fields of dots and crosses, before rescaling, will have a density which decays 
as $n\to \infty$. The rate of decay is not defined by the procedure. For example, in the KPZ
case it decays as $n^{-2/3}$. Later we will see cases where the density behaves as $n^{-1/2}$. In general the density will decay as $n^{-\alpha}$. A critical
exponent $\alpha \in (0,1)$ is a dots-crosses counterpart of the critical exponent $\xi$ introduced above to characterize large time displacements
of one-sided minimizers, i.e. $\alpha=\xi$. It is an interesting question whether values of $\alpha$ different from $2/3$ and $1/2$ can appear in physical problems.

\begin{conjecture}
For every critical exponent $\alpha\in(0,1)$, there is a fixed point for the renormalisation transformation $R$ with density scaling as  $n^{-\alpha}$.
The fixed points are stable except for one neutral direction corresponding to the 
parameter~$\alpha$.
It follows that if a random process has density decay $n^{-\alpha}$, it will converge to
the $\alpha$-fixed point under the action of $R$. In other words, the $\alpha$-fixed point
provides universal asymptotic statistics for all systems with exponent $\alpha$.
\end{conjecture}
In the above conjecture, the stability of fixed points is suggested when the point-connector configurations to which the renormalisation transformation is applied have fast
decay of space-time correlations. Otherwise, one cannot expect convergence to $\alpha$-fixed 
points. 

If the conjecture is correct, then it provides an insight on how far the KPZ universality extends.
What is actually required is monotonicity and the scaling $L\sim T^{2/3}$. The condition
of exact monotonicity can be probably replaced by a weaker assumption of approximate
monotonicity.

Let us stress again that the renormalisation transformation above deals only with
geometric data. In other words, we record only the positions of dots and crosses but not the values of action.
Thus it is very different from the renormalisation transformations based on so called
Airy sheet that have been discussed in the literature previously. The geometric objects like 
concentration dots and separation crosses that we use can be naturally defined for
minimisers and directed polymers. It is an interesting problem to find a counterpart
of these objects in the setting of random matrices.

\section{Coalescing Brownian Motions and Stochastic Flows}
\label{sec:colescing-BM}
There is one case where the renormalisation conjecture can be studied rigorously.
This is the case of the density scaling exponent $\alpha=1/2$. Then the configurations of
dots, crosses, and connectors, are independent for different strips.
Essentially this case was studied in \cite{Piterbarg-thesis},\cite{Piterbarg:MR1626162},\cite{2Piterbarg:MR1686850}. Below, we briefly discuss the setting and present the main results.

Consider the stochastic flow
\begin{equation}\label{eq:SF}
dx(t)=f_\omega (t,x(t))dW(t),
\end{equation}
where $W(t)$ is the standard Wiener process, and $f_\omega (t,x)$ is a random
function similar tho the random potential $F_\omega (t,x)$. We also assume that 
$f_\omega (t,x)$ space-time stationary,  smooth in $x$ and white in $t$.
The randomness of $f_\omega (t,x)$ and the one provided by the white noise are completely
separate, i.e., independent.
The stochastic flow (\ref{eq:SF}) is defined for a fixed realization of $f_\omega (t,x)$.
It corresponds to solving the SDE (\ref{eq:SF}) for all initial conditions $x(0)$ simultaneously
using the same white noise. 

It is well known that maps $S_W^t$ that assign solutions $x(t)$ to the initial conditions $x(0)$
 form  a family of diffeomorphisms almost surely in $W$. 
 These diffeomorphisms are monotone which also means that trajectories $x_1(t)$
and $x_2(t)$ for two different initial conditions cannot intersect. Also, it is easy to see that $x_2(t)-x_1(t)$ is a positive martingale
provided $x_2(0)>x_1(0)$. Hence there exist a $\lim_{t\to \infty}(x_2(t)-x_1(t)) = c(x_1,x_2)$.
 If the function
$f_\omega (t,x)$ is aperiodic, then $c(x_1,x_2)=0$ and all trajectories are asymptotic to
each other as $t \to \infty$.  On the other hand, if $f_\omega (t,x)$ is space-periodic, say,
$1$-periodic, then $c(x_1,x_2)$ can take any integer value.
In this case the diffeomorfism $S_W^t$  can be considered as a diffeomorfism of the
unit circle~$\T^1$. 

In the latter case, there is a strong similarity between the behaviour of one-sided minimisers
and trajectories of the stochastic flow (\ref{eq:SF}). 
Similarly to the case of Lagrangian minimisers, the family $(S_W^t)$ is hyperbolic.
There is also a unique random point similar to the global shock.
Namely, an exponentially small neighbourhood of this point gets expanded by $S_W^T$ onto almost the entire unit circle for large times $T$. At the same time, the rest of the unit circle contracts
into an exponentially small interval at time $T$. To find this random point of instability one has to take initial conditions at a time $T\gg 1$ and run the stochastic flow backwards in time until
time $t=0$. The concentration domain at time $t=0$ converges to the instability point in a limit
as $T\to \infty$. In a same way one can find a unique random point similar to the position of the unique global minimiser. In this case one has to put the initial conditions at time $-T$ and then run the stochastic flow forward in time until time $t=0$. In the limit as $T\to \infty$ the concentration 
interval shrinks to this special random point. The rate of exponential contraction and expansion
is determined by a non-random Lyapunov exponent. The behaviour is similar to 
the behaviour of one-sided minimisers in
the compact case (see Figure~\ref{fig:universal-cover}). The only difference is that the trajectories of the stochastic flow evolve in forward time.

The
situation changes in the non-compact setting. As we explained above, all trajectories are asymptotic to each other.
Now there are no points that are unstable for all times. However, for a given time scale $T$,
there are small intervals of instability corresponding to this time scale. Trajectories from these
exponentially small intervals are expanding to intervals of the length of order $\sqrt{T}$ at time $T$.
Nevertheless, if we wait for long enough time, these intervals will start to contract with
exponential rate. Again, the picture is similar to Figure~\ref{fig:clustering-of-minimisers} representing one-sided minimisers in the noncompact case. These intervals of instability are similar to old shocks of the age
of the order $T$ or larger. The only difference is the characteristic length scale for a large time $T$, which is $L\sim T^{1/2}$ now instead of $T^{2/3}$. Indeed, $T^{1/2}$ is a typical scale for the
diffusion process associated with the SDE (\ref{eq:SF}). This is exactly the difference between the
density scaling exponents which we discussed in the previous section. The behaviour is qualitatively 
similar. However the scaling limit of asymptotic statistics is different since it is determined by the
exponent $\alpha$.

Since the length scale is $T^{1/2}$, one should expect that two trajectories originating from distinct points $x$ and $y$ at distance $L$ apart will converge exponentially fast
but nonuniformly. Namely, $|x(t)-y(t)|\sim \exp(-\lambda (t -T(x,y))_+ )$, where $\lambda$ is a 
non-random Lyapunov exponent, and $T(x,y) \sim L^2$. This is also similar to the behaviour
of one-sided minimisers apart from the different scaling exponent. It was shown in \cite{Piterbarg-thesis},\cite{Piterbarg:MR1626162},\cite{2Piterbarg:MR1686850}
that in the diffusive rescaling, the stochastic flow converges to the coalescing Brownian motion.
However, the convergence has been proved only on a topological level. The hyperbolicity
picture presented above was not established rigorously, although we believe it can be done.

We next discuss the construction of the point fields and connectors. Essentially it is similar
to the construction in the case of one-sided minimisers. We first fix a large time scale $T\gg1$,
and consider the time strips $[iT, (i+1)T], \, i \geq 0$. Then for each strip we consider exponentially short intervals
of instability at time $iT$ which extend into long, of the order of $\sqrt{T}$, intervals at time $(i+1)T$,
and  exponentially short intervals of concentration of mass at time $(i+1)T$. We then apply
the diffusive scaling rescaling time by $T$ and space by  $const\,\sqrt{T}$. In the limit $T\to\infty$
we obtain for each rescaled time strip $[i,i+1]$ two point fields of density $1$: the field of crosses at time
$i$ and the field of dots at time $i+1$. Constant in the space scaling is needed only to ensure 
that the density of limiting point fields is equal to $1$. The connectors between two point fields in each strip
can be defined exactly in the same way as before. Indeed, the whole interval of points between
two crosses is mapped into a particular dot. Then by enumerating intervals between crosses
and dots we can define the connectors.
%$\xi_i$. 
We note that in the case of stochastic flows,
the point fields and connectors in different strips and independent, although two point fields
in a particular strip are still strongly correlated.  Hence, it is enough to describe statistics in a given
strip. Since the stochastic flow converges to the coalescing Brownian motion, the statistics
is completely determined by the latter, and, hence, is universal. 

Coalescing Brownian motion is a process defined by  a collection of independent Brownian motions
starting from every point $x \in \R^1$ at time $t=0$. It is also assumed that after two different Brownian motions meet at certain point of space-time they both continue to move together
in forward time. The system was studied by R.Arratia in the late 70s (\cite{Arratia:MR2630231}). It was shown 
that after any positive time~$t$ only countably many of distinct Brownian motions will be present,
and their locations at time $t$ form a locally finite point field of positive density. Now, let us consider
time $t=1$ and define a point field of positions of Brownian motions at this time (dots), and a point
field of crosses at time $t=0$. As usual each interval between crosses correspond to a particular 
dot at time 1. A connector field is also naturally defined. It follows immediately from
the scaling properties of the Brownian motion that these two point field and connector chosen independently for all the time strips $[i,i+1]$ form
a fixed point for the renormalisation scheme introduced in the previous section. Moreover,
it follows from \cite{Piterbarg-thesis},\cite{Piterbarg:MR1626162},\cite{2Piterbarg:MR1686850} that this fixed point is stable. 

We finish this section with another interesting property of the above fixed point.
It turns out that it carries certain exact integrability features. Namely, as it
was shown in \cite{2Piterbarg:MR1686850},
see also~\cite{Zaboronski:MR2259212},
\cite{Zaboronski:MR2851057} the correlation functions of the fixed point fields  are expressed by Pfaffians. More precisely, consider $2n$ ordered points on the real line
$x_1<x_2< \dots<x_{2n-1},x_{2n}$. They define $n$ intervals $I_1=[x_1,x_2], I_2=[x_3,x_4], \dots, I_n=[x_{2n-1}, x_{2n}]$. Consider now the event that all of these intervals are empty,
that is there are no crosses at time $t=0$ inside these intervals. Then the probability
of this event is given by the Pfaffian  of the following $2n\times 2n$ matrix $A$:
$a(i,i)=0, \, 1\leq i\leq 2n, \, a(i,j)=-a(j,i), \, i\neq j, a(i,j) = G(x_j-x_i), \, 1\leq i<j\leq 2n$, where the kernel $G(x)$ is defined by:
$$G(x) =1 - \frac{1}{\sqrt{4\pi t}}\int^x_{-x}\exp\left\{-\frac{y^2}{4t}\right\}dy.$$
It is natural to ask whether similar property holds for other fixed points corresponding to
different values of the parameter $\alpha$, in particular in the KPZ case.

\section{Renormalisation for Airy sheet}
\label{sec:renorm-airy}
In Section~\ref{sec:renorm-point}, we constructed a renormalisation scheme based on geometrical coding.
A more traditional
approach is based on so called
Airy processes. We define the Airy sheet in the following way. Consider a
problem of point-to-point minimisers on a large time interval $[0,T]$
and define the following  random process:
\begin{equation}\label{AI-1}
\bar A_{\omega,T}(x, y) = \frac{A_{\omega,T}(\mu(T) xT^{2/3}, \mu
(T) yT^{2/3}) - C(T)T}{\delta (T)T^{1/3}} -
\frac{\mu^2(T)}{2\delta(T)}(x-y)^2.
\end{equation}
Here $A_{\omega,T}(\mu (T)T^{2/3}x, \mu (T)T^{2/3}y)$ is the minimal
action between points
$\mu(T)T^{2/3}x$ and  $\mu (T)T^{2/3}y$ on the time interval $[0,T]$,
and the quadratic term subtracted above is the compensation  due to
the shape function $S_0(a)$ which is assumed to be quadratic:
$ S_0(a)=C+a^2/2$. We have seen above that this assumption is
satisfied for systems
in the case of quadratic Hamiltonians which satisfy the property of
shear invariance.
The constants $C(T), \mu (T)$, and $\delta(T)$ are determined by two
conditions. First of all
we choose $C(T)$ and $\delta(T)$ in such a way that the first two
moments of $\bar A_{\omega,T}(0, 0)$
match the first two moments of the GUE Tracy--Widom law. Then we choose
$\mu (T)= \sqrt{2\delta (T)}$, so that the coefficient in the
quadratic term in \eqref{AI-1} equals~1. Now, taking the limit
in distribution as $T\to \infty$, we obtain the Airy sheet $\Ai(x, y)$.
The constants
$C(T), \mu (T)$, and $\delta(T)$  will converge to non-universal
positive constants
$C, \mu, \delta$ where $C$ is the value of the shape function
at $a=0$, that is $C= S_0(0)$. Note that the Airy sheet
has not been constructed rigorously yet. However, the process $\Ai (0,
y)=\Ai_2(y)$ known as $\mathrm{Airy}_2$ process was rigorously constructed by
 M.Pr\"ahofer and H.Spohn (\cite{Prahofer-Spohn:MR1933446}, see also~\cite{Johansson:MR1737991}).  Airy
sheet
$\Ai(x, y)$ is translation-invariant on the $(x, y)$ plane. Its
one-point marginal distributions are given by the Tracy--Widom law for
the GUE ensemble of random matrices with the standard
normalisation. As we have seen above, this
requirement determines the choice of constants $C(T)$ and $\delta (T)$.

One can define the renormalization transformation $R$ associated to
doubling the length
of the time interval.
The idea is that we have two independent copies of the
same stationary process $A(x, z)$ and $A' (z, x)$ corresponding to two
time intervals $[0,T]$ and $[T,2T]$. Then a new stationary process
is defined by gluing these intervals (or the corresponding space-time
strips) together and minimising over a common variable $z$.
One also has to do rescaling since the time interval is of length $2T$
now. As before, we assume that
the marginal distributions of the process $A(x, z)$ share the first
two moments with the
GUE Tracy--Widom law. The total action between points $x$ and $z$ is given by
$A(x,z) + (x-z)^2$. Define now the following process:
\begin{equation}\label {RGAI}
B(x,y)= \min_{z}{[A(x,z) + A'(z,y) +(x-z)^2 +(z-y)^2]} -\frac{1}{2}(x-y)^2.
\end{equation}
One can show that by subtracting the last term we make $B(x,y)$ a
stationary process.
We can now define a new stationary process $R(A)$ which is the image
of the process~$A$ under the action of the renormalisation transformation $R$ acting in
the space of stationary processes. Namely, $R(A)(x,y)= \frac{1}{\delta
(A)} [B(\mu (A)x, \mu (A)y) - C(A)]$,
where $C(A), \, \delta (A)$ are determined by the requirement that the
first two moments
of $R(A)(0,0)$ are the same as before, that is the same as for the GUE
Tracy--Widom law.
The constant $\mu(A)=\sqrt{2\delta (A)}$. It is easy to see that up to
the shift and scaling the
total action between points $x,y$ on the double time interval is given
by $R(A)(x,y) +(x-y)^2$.
Notice that due to the  choice of $\mu(A)$ the coefficient in front of
the quadratic term is equal to 1.

The above formula can be considered as a renormalisation
transformation acting on the
space of translation-invariant random fields. It is believed that this
transformation has
a unique fixed point which is exactly the Airy sheet $\Ai(x,y)$. Moreover this fixed
point is stable in the space of translation-invariant random fields
with certain condition
on decay of correlations. Note that at the fixed point $A=\Ai(x,y)$ the
constants $C(A),\, \delta (A), \, \mu (A)$ are given by
$C(\Ai)=0, \, \delta (\Ai)=2^{1/3}, \mu(\Ai)= 2^{2/3}$.

The conjecture about the existence of a unique stable fixed point
obviously implies universality of the KPZ phenomenon. However at
present
there are no rigorous approaches to the analysis of this renormalisation
transformation.
A nice feature of (\ref {RGAI}) is that it is based on gluing
together two independent copies of the same process. However, the
analytical structure of the
right-hand side of (\ref {RGAI}) is very singular, which makes the
analysis of fixed points,
and, especially, their stability rather awkward.

Some connection between the Airy sheet process and point fields which
we considered
in this paper can be viewed in the following construction. Take a
realisation of the Airy process
and for a given $x$ define $y(x) = \argmin_{y} [\Ai(x, y) + (x -
y)^2]$. It turns out that the function $y(x)$ is locally constant. It
means that it does not change when $x$ changes locally. In other
words, there is a point field of points
$y$ where the minimum can be attained. This point field of special values of $y$ is
similar to the point field of dots that we discussed before.  The end-points of intervals 
for
different dots form a point field of crosses. Note that this
dots-crosses point field
is similar in spirit to the one constructed above for the one-sided
minimisers but they are not the same. In particular, the former cannot
be used for a simple geometrical renormalisation procedure.
It is tempting to glue together two
independent copies of the dots-crosses fields and define the
renormalisation procedure similarly to the construction in Section~\ref{sec:renorm-point}.
However, this will not be a
self-consistent renormalisation scheme.
Selection of the new dots-crosses field for the double strip will
require also information about
the realisation of $\Ai(x, y)$.
\section{Concluding remarks}
\label{sec:concluding}
In this paper, we presented arguments in support of several
conjectures. Some of them
were related to the existence of global solutions to the random
Hamilton--Jacobi equation and one-sided minimisers. The case of
positive viscosity corresponds to directed polymers. We discussed
why the results in both cases are supposed to be parallel. Indeed, in
the case of strong disorder,
directed polymers are localised and, hence, the situation is not much
different from the inviscid case of
one-sided minimisers. In the case of weak disorder, the mechanisms are
much simpler and rather
well understood. The above arguments apply in the case of quadratic
hamiltonians. In the general
case we define the process which can be considered as a
generalization of directed
polymers. We argue that the behaviour of the generalised directed
polymers must be similar
to the one in the quadratic case. Hence the above arguments can be
applied to the general case as
well.

We then concentrate on the 1D case and derive the KPZ scalings
exponents from the predicted
diffusive behaviour of the global solutions. After introducing the
system of point fields corresponding
to locations of concentration of one-sided minimisers and shocks
separating domains concentrated
at neighbouring locations, we define a renormalisation transformation
acting on the space of such systems. It is important that the
renormalisation transformation is defined in pure geometrical terms.
We predict that our renormalisation scheme has family
of fixed points parametrized by the scaling exponent $\alpha$ of
the decrease of density of point fields
with time. We also conjecture that every fixed point  is stable apart
from the obvious neutral direction corresponding to a change of the
parameter $\alpha$. Hence systems with the same scaling exponent will
have the same universal asymptotic statistical properties which are
determined by the corresponding fixed point. Based on the above
picture the following approach to the problem of KPZ universality
looks very natural. One has first to establish the scaling exponent
$\alpha=2/3$, and then prove stability of the fixed point for this
exponent. The main idea here is that
the exponent is determined by the physics of the problem, and then
monotonicity of interlacing
between minimisers and shocks requires particular statistical
behaviour. In other words, the situation is extremely rigid.
Basically, there exists only one statistics compatible with both
interlacing and monotonicity, and given scaling exponent $\alpha$.

Let us make a couple of remarks related to the Pfaffian property of
the dots-crosses fields.
We have seen that this property holds in the case  $\alpha=1/2$, and
suggested that this may
also be true for other values of $\alpha$. Notice that the kernel $G$
in the case $\alpha=1/2$
is simply a probability that two independent Wiener processes started
from two points distance $x$ apart will intersect until time 1. A
similar kernel can be defined in the general case. For the
dots-crosses field constructed for the Airy sheet the role of the
kernel $G$ would have been played by the kernel
$G_{KPZ}(x), $ which is the probability that $y(x) = \argmin_{y}
[\Ai(x, y) + (x - y)^2]= y(0) = \argmin_{y} [\Ai(0, y) + y^2]$.
However, apart from the analogy with the case $\alpha=1/2$ and the exact
integrability associated with the Airy sheet we don't have a serious
argument on why the dot-crosses field may satisfy the Pfaffian
property with this kernel.

Another remark concerns the limiting distribution for the
end-point of one-sided minimisers
or directed polymers  rescaled  by the factor $\mu T^{2/3}$. Consider
the one-sided minimiser
from the point $(0,0)$ and denote its position at time $-T$ by $\mu
(T) T^{2/3}y(0)$. The total action
consists of two parts corresponding to time intervals $[-T,0]$ and
$(-\infty,-T]$.
The first one is given by \[\delta(T) T^{1/3}[\bar A_{\omega, T}(0,y) +
y^2] + C(T)T].\] The second part is given by the global
solution $\Phi_\omega(-T, \mu (T)T^{2/3}y)$. Due to the diffusive
scaling it
can be presented as $\sigma \sqrt{\mu (T)} T^{1/3}W(y(0)) + \Phi_\omega(-T,
0)$, where $W(y)$ is the standard
Wiener process statistically independent from the Airy process $\Ai(x,
y)$. Adding two terms together and taking limit as $T \to \infty$
we get
$$y(0)= \argmin_{y}{\left[ \Ai_2(y) + \frac{\sigma \sqrt{\mu}}{\delta}W(y) + y^2\right]}.$$
Here $\mu, \delta$ are limiting values of $\mu (T), \delta (T)$, and $\sigma$ is
a normalizing constant in Conjecture~\ref{eq:CLT-conjecture}. The constants $\mu, \delta$, and
$\sigma$ are not
independent.  In fact, the coefficient $\frac{\sigma
\sqrt{\mu}}{\delta}W(y)$ is matching
the diffusion constant for the small increments of the $\Ai_2$ process
which are known to be diffusive
with the diffusion constant~2. Namely, $Var(\Ai_2(y)-\Ai_2(0))\sim 2|y|$
asymptotically as $y\to 0$.
Hence, $\frac{\sigma \sqrt{\mu}}{\delta}W(y)=\sqrt{2}W(y)$.
Substituting into the above formula
we get a universal probability distribution for the
rescaled end-point of the one-sided minimiser $\gamma_{\omega,0,0}(t),
t\in (-\infty, 0]$ corresponding  to space-time point $(0,0)$. Namely,
as $T\to \infty$
$$  \frac{\gamma_{\omega,0,0}(-T)}{\mu T^{2/3}} \to \argmin_{y}{\left[
\Ai_2(y) + \sqrt{2}W(y) + y^2\right]}.$$
The convergence above is in the distributional sense.
Notice that one can naturally define the asymptotic distribution for
the end-point for any
value of the critical exponent $\alpha$. It is enough to run the
renormalisation scheme until
the strip of time length $n$ is reached. Then we take the concentration
dot corresponding
to the interval between two crosses containing the origin, rescale its
position by $n^\alpha$,
and take limit as $n\to \infty$.

We finish with two questions which we find interesting and challenging.
We have seen before that dots-crosses point fields can be constructed
starting from the Airy
sheet process. Going in other direction, is it possible to reconstruct
Airy processes starting
from the dots-crosses point fields?

Another question  is related to the scaling parameter $\alpha$ which
determines different fixed
points of the geometrical renormalization transformation. We have seen
that $\alpha=2/3$
corresponds to the KPZ phenomenon, while $\alpha=1/2$ is connected
with the stochastic flows
of diffeomorphisms. One can ask whether other values of $\alpha$ can appear in
some natural physical or mathematical settings. It is also interesting
whether one can define an
analogue of the Airy processes for the values of $\alpha$ different from $2/3$.

%\appendix
\section{Appendix. Generalized polymers in the quadratic Hamiltonian case}
\label{sec:equiv-of-poly-for-Burgers}
The goal of this section is to explain that in the case  of quadratic
Hamiltonian, the generalized polymers that we introduced in 
Section~\ref{sec:HJ-polymers} coincide with 
the usual polymers defined  via  Gibbs modifications of the Wiener measure.

We first explain that polymer measures arising in the Hopf--Cole--Feynman--Kac solution of the Burgers-- Hamilton--Jacobi  equation define diffusion processes with drifts given by the velocity field solving the Burgers equation. Then we recall a general variational
principle for Gibbs distributions and use it to derive a variational characterization of the Burgers polymer measures, thus showing that the Hopf--Cole--Feynman--Kac polymer measures coincide with generalized polymer measures constructed  via  stochastic control.  Our results are similar to existing variational characterizations of  Wiener functionals in
\cite{Boue-Dupuis:MR1675051},\cite{Ustunel:MR3255483},\cite{Eyink-Drivas:MR3299883}, see
also~\cite{Chen-Georgiou-Pavon:MR3489825}, but present a different point of view.

Throughout this section we fix $\visc>0$.

\subsection{Burgers polymers as diffusions}

The Feynman--Kac formula~\eqref{eq:Feynman-Kac} solving the linear heat equation~\eqref{eq:parabolic-model}
can be naturally interpreted as integration with respect to a Gibbs measure on paths constructed using the Wiener
measure as the free measure and Boltzmann weight composed of contributions from the initial condition
and the potential accumulated by paths. Let us define these objects more precisely.

For every $x\in\R^d$ and $t>0$, we define a version $\Ww_{t,x}$ of the Wiener measure to be the distribution of~$\gamma_s=x+\sqrt{2\visc}B_s$ on $C([0,t])=C([0,t],\R^d)$, where $B_s$ is a standard
Wiener process in $\R^d$. We define the energy function on paths by
\begin{equation}
\label{eq:Energy}
 E(\gamma)=\frac{1}{2\visc}\left[-\int_0^{t}F(t-s,\gamma_s)ds+\Phi_0(\gamma_t)\right],\quad \gamma\in C([0,t]),
\end{equation}
where we used a fixed realization of smooth potential $F$.
Then the polymer measure $\Pp_{t,x}$ associated to $F$ is also a measure on $C([0,t])$,
 absolutely continuous with respect to $\Ww_{t,x}$, with density 
\begin{equation}
\label{eq:polymer-density}
\frac{d \Pp_{t,x}}{d\Ww_{t,x}}(\gamma)=\frac{e^{-E(\gamma)}}{Z[Z_0](t,x)},\quad \gamma\in C([0,t]),
\end{equation}
where $Z_0=\exp\{-\Phi_0/(2\nu)\}$, and the partition function $Z[Z_0](t,x)$ is given by the Feynman--Kac formula
\eqref{eq:Feynman-Kac} that can be rewritten as
\[
Z[Z_0](t,x)=\int_{ C([0,t],\R^d)}e^{-E(\gamma)}\Ww_{t,x}(d\gamma).
\]
For an arbitrary Markov control $v:[0,t]\times\R^d\to\R^d$, we denote by $\Qq_{t,x,v}$ the distribution
of solutions of the following SDE on $C[0,t]$:
\begin{align}
\label{eq:controlled-diffusion} 
d\gamma_s&=-v(t-s,\gamma_s)ds+\sqrt{2\visc}dB_s,\\
 \gamma_0&=x,
 \label{eq:initial-condition-for-controlled-diffusion}
\end{align}
which is simply the forward version of the SDE in \eqref{eq:controlled-sde}--\eqref{eq:controlled-sde-initial}.
\begin{theorem} \label{thm:Gibbs-coincides-with-u-controlled}
The distribution $\Pp_{t,x}$ of paths $\gamma$ coincides with $\Qq_{t,x,u}$,
where $u(\cdot,\cdot)$ is the solution  of the Burgers equation~\eqref{eq:Burgers}.
\end{theorem}
\bpf
Recalling that due to \eqref{eq:Hopf-Cole-0},
\[
u(s,y)=\nabla\Phi(s,y)=-2\visc\nabla \ln Z(s,y)=-2\visc\frac{\nabla Z(s,y)}{Z(s,y)}, 
\]
we use Girsanov's theorem to write
\begin{align*}
\frac{d \Qq_{t,x,u}}{d \Ww_{t,x}}(\gamma)&=\exp\left\{\frac{1}{2\visc}\left[\int_0^t -( u(t-s,\gamma_s)\cdot d\gamma_s) -\frac{1}{2}\int_0^t |u(t-s,\gamma_s)|^2 ds\right]\right\} 
\\&=\exp\left\{\int_0^t \left( \frac{\nabla Z(t-s,\gamma_s)}{ Z(t-s,\gamma_s)}\cdot d\gamma_s\right) -\visc\int_0^t  \frac{|\nabla Z(t-s,\gamma_s)|^2}{ Z^2(t-s,\gamma_s)} ds\right\}. 
\end{align*}
Here $\gamma$ is a Brownian motion under $\Ww_{t,x}$, with  $d\langle \gamma^i,\gamma^j\rangle_s=2\visc \delta^{ij}ds$ and $\gamma_0=x$. The integral with respect to $d\gamma_s$ is an It\^o integral. 
Comparing this to~\eqref{eq:polymer-density}, we only need to check that $X_t=-\ln Z(t,x)$,
where for $r\in[0,t]$, the process $X_r$ is defined by
\begin{multline}
\label{eq:X-in-proof-of-Girsanov-FK}
X_r= \int_0^r \left( \frac{\nabla Z(t-s,\gamma_s)}{ Z(t-s,\gamma_s)}\cdot d\gamma_s\right) -\visc\int_0^r  \frac{|\nabla Z(t-s,\gamma_s)|^2}{ Z^2(t-s,\gamma_s)} ds
\\+\frac{1}{2\visc}\left[-\int_0^{r}F(t-s,\gamma_s)ds+\Phi(t-r,\gamma_r)\right].
\end{multline}
Since due to~\eqref{eq:Hopf-Cole-0},
\begin{equation}
\frac{1}{2\visc}\Phi(t-r,\gamma_r)= -\ln Z(t-r,\gamma_r),\quad r\in [0,t],
\end{equation}
we have $X_0=-\ln Z(t,x)$, and 
it suffices to prove that $dX_r=0$. The rest of the proof is a standard stochastic calculus calculation. By the It\^o formula, we have
\begin{multline*}
 d\left[\frac{1}{2\visc}\Phi(t-r,\gamma_r)\right]= -d\ln Z(t-r,\gamma_r)
 \\=
 -\frac{\bigl(-\partial_tZ(t-r,\gamma_r)    +   \visc \Delta Z(t-r,\gamma_r)            \bigr)dr + (\nabla Z(t-r,\gamma_r)\cdot d\gamma_r)}{Z(t-r,\gamma_r)} \\
                           +\frac{\visc |\nabla Z(t-s,\gamma_s)|^2}{ Z^2(t-r,\gamma_r)}dr.
\end{multline*}
Plugging this into~\eqref{eq:X-in-proof-of-Girsanov-FK} and using equation~\eqref{eq:parabolic-model}, we obtain $dX_r=0$, which completes the proof. \epf

\subsection{The variational principle for Gibbs measures}
The following is a standard variational characterization of Gibbs measures
that is sometimes called Gibbs inequality for Kullback--Leibler divergence. 
\begin{theorem}\label{lem:general-var-principle} 
Let $\mu$ be a probability measure on a measurable space $(\X,\Xc)$ and let $E:\X\to\R$ be a measurable function. Let 
$ \Delta$ be the space of measurable nonnegative functions $p:\X\to\R_+$ satisfying $\int_{\X}\mu(dx)p(x)=1$.
We introduce the average energy functional
\[
 I(p)=\int \mu(dx) p(x)E(x),\quad p\in\Delta,
\]
the entropy functional
\[
 h(p)= -\int_{\X}\mu(dx) p(x)\ln p(x),\quad p\in\Delta,
\]
and the free energy functional
\[
 G(p)=I(p)-h(p),\quad p\in\Delta.
\]
Then the minimum of the free energy is uniquely provided by the Gibbs measure with free measure $\mu$ and energy $E$. Namely,
introducing the the partition function~$Z$ by
\[
 Z=\int_{\X}\mu(dx)e^{-E(x)}
\]
and the Gibbs density $q\in\Delta$ by
\[
 q(x)=\frac{e^{-E(x)}}{Z},\quad x\in\X,
\]
we have
\[
 \inf_{p\in\Delta} G(p)=G(q)=-\ln Z.
\]
\end{theorem}
\bpf We provide the well-known proof here for completeness since it is very short. Let $\varphi(x)=x\ln x$ for $x>0$ and $\varphi(0)=0$. 
For every $p\in\Delta$,
we have
\begin{align*}
I(p)-h(p)+\ln Z&= \int_{\X}\mu(dx)p(x)E(x)+\int_{\X}\mu(dx)p(x)\ln p(x)+\ln Z 
\\&= \int_{\X}\mu(dx)p(x)\ln (p(x) e^{E(x)}Z) =  \int_{\X}\mu(dx)p(x)\ln \frac{p(x)}{q(x)}
\\&=\int_{\X}\mu(dx)q(x)\frac{p(x)}{q(x)}\ln \frac{p(x)}{q(x)}=\int_{\X}\mu(dx)q(x)\varphi\left(\frac{p(x)}{q(x)}\right)
\\&\ge \varphi\left( \int_{\X}\mu(dx)q(x)\frac{p(x)}{q(x)}\right)=\varphi(1)=0,
\end{align*}
where we used convexity of $\varphi$ and Jensen's inequality.
Also, if $p=q$, then this inequality becomes an identity, and the theorem follows,
since Jensen's inequality is strict unless $p(x)/q(x)=1$ for $\mu$-a.e.\ $x\in\X$.
\epf

\subsection{Stochastic control characterization of the Burgers polymer}
Let us state the main result of this section. Let us fix $\nu>0$, $t>0$, and the initial condition $\Phi_0$.
\begin{theorem}\label{thm:FK-same-as-control} There exists a unique Markov control $v:[0,t]\times \R^d\to\R^d$ realizing the minimum of the
functional 
\[
G(v)=\frac{1}{2\visc}\E_{\Qq_{t,x,v}} \left[\frac{1}{2} \int_0^t v^2(t-s,\gamma_s)ds-\int_0^t F(t-s,\gamma_s)ds+\Phi_0(\gamma_t)\right].
\]
This optimal control coincides with the solution $u$ of the Burgers equation~\eqref{eq:Burgers}. Moreover, $\Qq_{t,x,u}$ is the polymer measure: it coincides with $\Pp_{t,x}$
defined in~\eqref{eq:polymer-density}. 
\end{theorem}

\bpf
We are going to apply Theorem~\ref{lem:general-var-principle} to $\mu=\Ww_{t,x}$ and energy $E$ defined in~\eqref{eq:Energy}.  Taking any
 control  
 $v:[0,t]\times\R^d\to\R^d$.  
and denoting by $p_v$ the Girsanov density
$\frac{d \Qq_{t,x,v}}{d \Ww_{t,x}}$, we see that, in the notation of Theorem~\ref{lem:general-var-principle}, 
\[
 I(p_v)=\int \Qq_{t,x,v}(d\gamma) E(\gamma)= \frac{1}{2\visc}\E_{\Qq_{t,x,v}}\left[-\int_0^{t}F(t-s,\gamma_s)ds+\Phi_0(\gamma_t)\right], 
\]
so if we show that 
\begin{equation}
\label{eq:Girsanov-entropy}
  h(p_v)=-\frac{1}{2}\cdot\frac{1}{2\visc} \E_{\Qq_{t,x,v}} \int_0^t\int_0^t |v(t-r,\gamma_r)|^2 dr,
\end{equation}
then Theorem~\ref{lem:general-var-principle} will imply that the values of $G(v)$ are bounded below by
the free energy of 
the Gibbs measure associated with energy $E(\cdot)$ and free measure $\Ww_{t,x}$. According to Theorem~\ref{thm:Gibbs-coincides-with-u-controlled} that Gibbs measure coincides with $\Qq_{t,x,u}$, so our
claims follow.

Let us check~\eqref{eq:Girsanov-entropy}. We note that $p_v=e^{X_t}$, where 
\[
 X_r(\gamma)=\frac{1}{2\visc}\left[-\int_0^r \bigl(v(t-s,\gamma_s)\cdot d\gamma_s\bigr)
 -\frac{1}{2}\int_0^r |v(t-s,\gamma_s)|^2ds\right],\quad r\in[0,t].
\]
Basic stochastic calculus implies:
\begin{align*}
  d e^{X_r}&= e^{X_r}dX_r+\frac{1}{2}e^{X_r}d[X]_r
          \\&=-\frac{e^{X_r}}{2\visc}\left[\bigl(v(t-r,\gamma_r)\cdot d\gamma_r\bigr)+\frac{1}{2} |v(t-r,\gamma_r)|^2dr\right]
+\frac{1}{2}e^{X_r}\cdot\frac{1}{2\visc} |v(t-r,\gamma_r)|^2dr\\ &
=-\frac{e^{X_r}}{2\visc}\bigl(v(t-r,\gamma_r)\cdot d\gamma_r\bigr).
\end{align*}

\begin{flalign*}
 d(X_re^{X_r})=&X_r d e^{X_r}+e^{X_r}dX_r+d[X, e^X]_r&
\\=& - \frac{X_re^{X_r}}{2\visc}\bigl(v(t-r,\gamma_r)\cdot d\gamma_r\bigr) 
\\&- \frac{e^{X_r}}{2\visc}\left[\bigl(v(t-r,\gamma_r)\cdot d\gamma_r\bigr)+\frac{1}{2} |v(t-r,\gamma_r)|^2 dr\right]+\frac{e^{X_r}}{2\visc}|v(t-r,\gamma_r)|^2 dr&
\\ =& \frac{1}{2}\cdot \frac{e^{X_r}}{2\visc}|v(t-r,\gamma_r)|^2 dr+dM_t,&
\end{flalign*}
where $M$ is a local martingale. Treating it as a martingale (in general, a more
delicate argument is needed), we obtain
\begin{multline}
\label{eq:entropy-aux}
 h(p_v)= -\E_{\Ww_{t,x}}[ X_t e^{X_t}]=-\E_{\Ww_{t,x}}\left[\frac{1}{2} \int_0^t 
\frac{e^{X_r}}{2\visc}|v(t-r,\gamma_r)|^2 dr \right] \\
=-\frac{1}{2} \int_0^t \E_{\Ww_{t,x}} \left[\frac{e^{X_r}}{2\visc}|v(t-r,\gamma_r)|^2\right] dr. 
\end{multline}
The martingale property of the Girsanov density process $(e^{X_r})_{r\in[0,t]}$  with respect to the natural filtration $(\Fc_r)_{r\in[0,t]}$ of $\gamma$ implies
\begin{align*}
 \E_{\Ww_{t,x}} [e^{X_t}|v(t-r,\gamma_r)|^2]&= \E_{\Ww_{t,x}}\E_{\Ww_{t,x}} [e^{X_t}|v(t-r,\gamma_r)|^2|\Fc_r]
\\&=\E_{\Ww_{t,x}}[|v(t-r,\gamma_r)|^2\E_{\Ww_{t,x}} 
[e^{X_t} |\Fc_r]]
\\&=\E_{\Ww_{t,x}}[|v(t-r,\gamma_r)|^2e^{X_r}],
\end{align*}
so plugging this into~\eqref{eq:entropy-aux}, we obtain
\begin{align*}
 h(p_v)&=-\frac{1}{2} \cdot\frac{1}{2\visc} \int_0^t \E_{\Ww_{t,x}} \left[e^{X_t}|v(t-r,\gamma_r)|^2\right] dr
\\&=-\frac{1}{2}\cdot\frac{1}{2\visc} \E_{\Ww_{t,x}} \left[e^{X_t}\int_0^t |v(t-r,\gamma_r)|^2 dr \right]
\\&=-\frac{1}{2}\cdot\frac{1}{2\visc} \E_{\Qq_{t,x,v}} \int_0^t\int_0^t |v(t-r,\gamma_r)|^2 dr,
\end{align*}
which completes the derivation of~\eqref{eq:Girsanov-entropy} and the entire proof.
\epf

\bibliographystyle{plain-initials}
\bibliography{Burgers}

\end{document}